\documentclass[journal]{IEEEtran}
\usepackage{amsmath,epsfig,amssymb}
\usepackage{epsfig,subfigure}
\usepackage{graphicx}
\usepackage{algorithm,algorithmic}
\usepackage{stackengine}
\usepackage{cite}
\usepackage{url}
\usepackage{color}
\usepackage{bm}

\def\cG{\mathcal{G}}

\def\D{\mathbf{D}}
\def\I{\mathbf{I}}
\def\L{\mathbf{L}}

\def\I{\mathbf{I}}

\def\W{\mathbf{W}}

\def\x{\mathbf{x}}
\def\y{\mathbf{y}}

\DeclareMathOperator*{\argmin}{arg\,min}

\begin{document}

%
\title{Graph Spectral Image Processing}
%
%
%

\author{Gene~Cheung,~\IEEEmembership{Senior Member,~IEEE,}
        Enrico~Magli,~\IEEEmembership{Fellow,~IEEE,}
        Yuichi~Tanaka,~\IEEEmembership{Member,~IEEE,}
        and~Michael~Ng,~\IEEEmembership{Senior Member,~IEEE}
\thanks{Manuscript received August 30, 2017; revised November 30, 2017.}
\thanks{Y. Tanaka was supported in part by JSPS KAKENHI under Grant Number JP16H04362 and JST PRESTO under Grant Number JPMJPR1656. M. Ng was supported in part by HKRGC GRF 12302715, 12306616 and 12200317, and HKBU RC-ICRS/16-17/03.}}

%
%

\markboth{submitted to Proceedings of the IEEE}%
{Shell \MakeLowercase{\textit{et al.}}: Bare Demo of IEEEtran.cls for IEEE Journals}
%



\maketitle

\begin{abstract}
Recent advent of graph signal processing (GSP) has spurred intensive studies of signals that live naturally on irregular data kernels described by graphs (\textit{e.g.}, social networks, wireless sensor networks).
Though a digital image contains pixels that reside on a regularly sampled 2D grid, if one can design an appropriate underlying graph connecting pixels with weights that reflect the image structure, then one can interpret the image (or image patch) as a signal on a graph, and apply GSP tools for processing and analysis of the signal in graph spectral domain.
In this article, we overview recent graph spectral techniques in GSP specifically for image / video processing.
The topics covered include image compression, image restoration, image filtering and image segmentation. 
\end{abstract}


%
\IEEEpeerreviewmaketitle

\section{Introduction}

\textit{Graph signal processing} (GSP) is the studies of signals that live on irregularly structured data kernels described by graphs \cite{shuman13}, such as social networks and wireless sensor networks.
The underlying graph typically reveals signal structures; an edge $(i,j)$ with large weight $w_{i,j}$ connecting nodes $i$ and $j$ means that the signal samples at $i$ and $j$ are expected to be similar or correlated.
Though a digital image contains pixels that reside on a regularly sampled 2D grid, one can nonetheless interpret an image (or an image patch) as a signal on a graph, with edges that connect each pixel to its neighborhood of pixels. 
By choosing an appropriate graph that reflects the intrinsic image structure, a spectrum of graph frequencies can be defined through eigen-decomposition of the graph Laplacian matrix \cite{chung97}, and notions like transforms \cite{shen10pcs,hu12icip,zhang13,hu2015b,hu2015,fracastoro2017}, wavelets \cite{ram2011,narang2012,narang2013}, smoothness \cite{chen15,chen14,pang14,pang15,pang17} etc can be correspondingly derived.
Then a target image (or image patch) can be decomposed and analyzed spectrally on the chosen graph using developed GSP tools---analogous to frequency decomposition of square pixel blocks via known transforms like discrete cosine transform (DCT).
Recently, this graph spectral interpretation of traditional 2D images has led to new insights and understanding, resulting in optimization of both the underlying graph and the graph-based processing tools that shows demonstrable gain in a number of traditional image processing areas, including image compression, restoration, filtering and segmentation\footnote{We note that while graph has been used extensively as an abstraction for image processing in the past \cite{lezoray17}, we focus in this article in particular recent developed techniques that process or analyze image signals in appropriately chosen graph spectral domains.}.

For image compression, a Fourier-like transform for graph-signals called \textit{graph Fourier transform} (GFT) \cite{shuman13} and many variants \cite{shen2010,hu2012,zhang13,hu2015b,hu2015,fracastoro2017} have been used as adaptive transforms for coding of piecewise smooth (PWS) and natural images. 
Because the underlying graph used to define GFT can be different for each code block, the cost of describing the graph as well as the cost of coding GFT coefficients to represent the signal must both be taken into consideration.
For wavelets on graphs \cite{ram2011,narang2012,narang2013}, where the conventional notion of ``downsampling by 2" is ill-defined for irregular data kernels, how to define critically sampled perfect reconstruction filterbanks (with (bi)orthogonal conditions) using appropriate downsamplers has been a challenge. 
We review proposals in designs of graph transforms and wavelets for image / video compression in Section\;\ref{sec:compress}. 

For image restoration such as denoising and deblurring, how to design appropriate signal priors to regularize otherwise ill-posed problems is a major challenge. 
Notions of sparsity \cite{hu13} and signal smoothness \cite{belkin04,chen14,pang17,berger17} can also be generalized to the graph-signal domain. 
Wiener filtering for graph-signals, which first requires a proper definition of wide sense stationarity for irregular graph data kernels, was recently developed \cite{perraudin17}.
We review popular graph-based restoration techniques in Section\;\ref{sec:restore}.

Spectral filtering is a fundamental image processing operation. 
It turns out that the well-known bilateral filter for image denoising \cite{tomasi98} can be interpreted as a linear low-pass filter for a specific graph \cite{gadde13}. 
Other diffusion and edge-preserving smoothing operators are also discussed in Section\;\ref{sec:filter}. 
Popular applications such as image retargeting and non-photorealistic rendering of images are also overviewed. 
Finally, fast implementation of graph filters using Chebyshev polynomial approximation is discussed.

Image segmentation is an old computer vision problem, and there is a long history of graph-based approaches such as graph cuts \cite{shi00,boykov01}. 
More recent models such as the Mumford-Shah model \cite{boykov06} and graph biLaplacian \cite{fang2017} are discussed in Section\;\ref{sec:segment}.


\section{Preliminaries}
\label{sec:prelim}
\subsection{Graph Definition}
\label{subsec:defn}

We first introduce common definitions and concepts in GSP for use in later sections. 
A graph $\mathcal{G}(\mathcal{V}, \mathcal{E}, \mathbf{W})$ contains a set $\mathcal{V}$ of $N$ nodes and a set $\mathcal{E}$ of $M$ edges. 
Each existing edge $(i,j) \in \mathcal{E}$ is undirected and contains an edge weight $w_{i,j}$, which is typically positive; a large positive $w_{i,j}$ would mean that samples at nodes $i$ and $j$ are expected to be similar / correlated.
Common for images, weight $w_{i,j}$ of an edge connecting nodes (pixels) $i$ and $j$ is computed using a Gaussian kernel, as done in the bilateral filter \cite{tomasi98}:
\begin{equation}
w_{i,j} = \exp \left( - \frac{\|\mathbf{l}_i - \mathbf{l}_j\|_2^2}{\sigma_l^2} \right) 
\exp \left( - \frac{\|x_i - x_j\|_2^2}{\sigma_x^2} \right) 
\label{eq:edgeWeight}
\end{equation}
where $\mathbf{l}_i$ is the location of pixel $i$ on the 2D image grid, $x_i$ is the intensity of pixel $i$, and $\sigma_l^2$ and $\sigma_x^2$ are two parameters.
Hence $0 \leq w_{i,j} \leq 1$. Larger geometric and/or photometric distances between pixels $i$ and $j$ would mean a smaller weight $w_{i,j}$.
Edge weights can alternatively be defined based on local pixel patches, features, etc \cite{milanfar13}. 
To a large extent, the appropriate definition of edge weight (inter-node similarity) is application-dependent; we will introduce various definitions for different applications in the sequel.

More generally, a suitable graph can be constructed from a machine learning perspective---given multiple signal observations, identify a graph structure that best fits the observed data given a fitting criterion or model assumptions \cite{friedman08,daitch09,cheng10,egilmez17}.
For example, \textit{graphical lasso} in \cite{friedman08} computes a sparse inverse covariance matrix (precision matrix) assuming a Gaussian Markov Random Field (GMRF) model and a sparse graph. 
Graph learning is a fundamental problem in GSP and is discussed extensively in another article in this special issue. 

A \textit{graph-signal} $\x$ on $\cG$ is a discrete signal of dimension $N$---one sample $x_i \in \mathbb{R}$ for each node\footnote{If a graph node represents a pixel in an image, each pixel would typically have three color components. For simplicity, one can treat each color component separately as a different graph-signal.} $i$ in $\mathcal{V}$.
Assuming nodes are appropriately labeled from $1$ to $N$, we can treat a graph-signal simply as a vector $\x \in \mathbb{R}^N$.

\subsection{Graph Spectrum}
\label{subsec:spectrum}

Given an edge weight matrix $\mathbf{W}$ where $W_{i,j} = w_{i,j}$, we define a diagonal \textit{degree matrix} $\mathbf{D}$, where $d_{i,i} = \sum_{j} W_{i,j}$. 
A \textit{combinatorial graph Laplacian matrix} $\mathbf{L}$ is $\mathbf{L} = \mathbf{D} - \mathbf{W}$ \cite{shuman13}.
Because $\mathbf{L}$ is symmetric, one can show via the Spectral Theorem that it can be eigen-decomposed into:
\begin{equation}
\label{eq:eigendecomposition}
\mathbf{L} = \mathbf{U} \mathbf{\Lambda} \mathbf{U}^T
\end{equation}
where $\mathbf{\Lambda}$ is a diagonal matrix containing real eigenvalues $\lambda_k$ along the diagonal, and $\mathbf{U}$ is an eigen-matrix composed of orthogonal eigenvectors $\mathbf{u}_i$ as columns. 
If all edge weights $w_{i,j}$ are restricted to be positive, then graph Laplacian $\mathbf{L}$ can be proved to be \textit{positive semi-definite} (PSD)\cite{chung97}\footnote{One can prove that a graph $\mathcal{G}$ with positive edge weights has PSD graph Laplacian $\mathbf{L}$ via the Gershgorin circle theorem: each Gershgorin disc corresponding to a row in $\mathbf{L}$ is located in the non-negative half-space, and since all eigenvalues reside inside the union of all discs, they are non-negative.}, meaning that $\lambda_k \geq 0, \forall k$ and $\mathbf{x}^T \mathbf{L} \mathbf{x} \geq 0, ~\forall \mathbf{x}$. 
Non-negative eigenvalues $\lambda_k$ can be interpreted as \textit{graph frequencies}, and eigenvectors $\mathbf{U}$ interpreted as corresponding graph frequency components. Together they define the \textit{graph spectrum} for graph $\mathcal{G}$.

The set of eigenvectors $\mathbf{U}$ for $\mathbf{L}$ collectively form the \textit{graph Fourier Transform} (GFT) \cite{shuman13}, which can be used to decompose a graph-signal $\x$ into its frequency components via $\boldsymbol{\alpha} = \mathbf{U}^T \mathbf{x}$, similar to known discrete transforms such as DCT. 
In fact, one can interpret GFT as a generalization of known transforms like DCT; 
see Shuman et al. \cite{shuman13} for details.

Note that if the multiplicity $m_k$ of an eigenvalue $\lambda_k$ is larger than 1, then the set of eigenvectors that span the corresponding eigen-subspace of dimension $m_k$ is non-unique. 
In this case it is necessary to specify the graph spectrum as the collection of eigenvectors $\mathbf{U}$ themselves.
See more discussion on this issue in the compression context in Section\;\ref{sec:compress}.

If we consider also negative edge weights $w_{i,j}$ that reflect inter-pixel dissimilarity / anti-correlation, then graph Laplacian $\L$ can be indefinite.
We discuss a few recent works that employ negative edges in Section\;\ref{sec:restore}.

\subsection{Variation Operators}
\label{subsec:variation}

Closely related to the combinatorial graph Laplacian $\L$ are other variants of Laplacian operators, each with its own unique spectral properties.
A \textit{normalized graph Laplacian} $\L_n = \D^{-1/2} \L \D^{-1/2}$ is a symmetric normalized variant of $\L$.
In contrast, a \textit{random walk graph Laplacian} $\L_r = \D^{-1} \L$ is an asymmetric normalized variant of $\L$.  
A \textit{generalized graph Laplacian} $\L_g = \L + \mathrm{diag}(d_{i,i})$ is a graph Laplacian with self-loops $d_{i,i}$ at nodes $i$---called \textit{loopy graph Laplacian} in \cite{dorfler13}---resulting in a general symmetric matrix with non-positive off-diagonal entries \cite{biyikoglu05}.
Eigen-decomposition can also be performed on these operators to acquire a set of graph frequencies and frequency components. 
For example, normalized variants $\L_n$ and $\L_r$ share the same eigenvalues between $0$ and $2$.
While $\L$ and $\L_n$ are both symmetric, $\L_n$ does not have the constant vector as an eigenvector. 
Asymmetric $\L_r$ can be symmetrized via left and right diagonal matrix multiplications \cite{milanfar13_siam}.
We will discuss different choices of variation operators in the sequel for different applications\footnote{In \cite{ramano17}, a general denoising regularization term is proposed where the penalty is proportional to the inner product between the signal $\x$ and its denoised residual $\x - f(\x)$; $\x^T \L \x$ being an example if $\I - \L$ is interpreted as a denoising operator. The main goal of \cite{ramano17} is to show this regularization term can be used as an engine for more general inverse problems, similar to \textit{plug-and-play priors} ($P^3$) \cite{venka13}. In contrast, our goal here is to show different graph variation operators have different characteristics that are suitable for different applications.  }.

\subsection{Graph-Signal Priors}
\label{subsec:smooth}

Traditionally, for graph $\cG$ with positive edge weights, signal $\x$ is considered \textit{smooth} if each sample $x_i$ on node $i$ is similar to samples $x_j$ on neighboring nodes $j$ with large $w_{i,j}$.  
In the graph frequency domain, it means that $\x$ contains mostly low graph frequency components; \textit{i.e.}, coefficients $\boldsymbol{\alpha} = \mathbf{U}^T \x$ are zeros for high frequencies.
The smoothest signal is the constant vector---the first eigenvector $\mathbf{u}_1$ for $\mathbf{L}$ corresponding to the smallest eigenvalue $\lambda_1 = 0$.

Mathematically, we can write that a signal $\x$ is smooth if its \textit{graph Laplacian regularizer} $\x^T \mathbf{L} \x$ is small \cite{pang14,pang15,pang17}. 
Graph Laplacian regularizer can be expressed as:
\begin{align}
\mathbf{x}^T \mathbf{L} \mathbf{x}  & = \sum_{(i,j) \in \mathcal{E}} w_{i,j} \left( x_i - x_j \right)^2 = \sum_{k} \lambda_k \, \alpha_k^2
\label{eq:smoothness}
\end{align}
Because $\mathbf{L}$ is PSD, $\x^T\mathbf{L}\x$ is lower-bounded by $0$, achieved when $\x = c \mathbf{u}_1$ for some scalar constant $c$.

In \cite{chen15} the adjacency matrix $\W$ is interpreted as a shift operator, and thus graph-signal smoothness is defined instead as the difference between a smooth signal $\x$ and its shifted version $\W \x$. 
Specifically, graph total variation based on $l_p$-norm is:
\begin{align}
\mathrm{TV}_{\mathbf{W}}(\mathbf{x}) = \left\| \x - \frac{1}{|\lambda_{\max}|} \W \x \right\|_p^p
\label{eq:smoothness2}
\end{align}
where $p$ is a chosen integer. 
More specifically, a quadratic smoothness prior is defined in \cite{chen14} (also in \cite{ramano17}):
\begin{equation}
S_2(\mathbf{x}) = \frac{1}{2} \| \mathbf{x} - \mathbf{W} \mathbf{x} \|_2^2
\label{eq:smoothness3}
\end{equation}

Besides smoothness, sparsity of graph-signals with respect to a trained graph dictionary can also be used as a prior \cite{thanou15}.
Specifically, to effectively represent signals on different graph topologies, graph atoms are constructed as polynomials of the graph Laplacian.
Preliminary results in \cite{thanou15} show its potential, but we will not discuss this further in the sequel.


\section{Graph-based Image Compression}
\label{sec:compress}
Image compression refers to the process of encoding an image $\mathbf{x}$ onto a codeword $c(\mathbf{x})$, minimizing distortion in the reconstructed image $\widehat{\mathbf{x}}$ for a given target bit-rate $R_T$, i.e.
\begin{equation}
\min D(\mathbf{x},\widehat{\mathbf{x}}) \; \; {\rm subject} \; {\rm to} \; \; R(c(\mathbf{x})) \leq R_T,
\label{eq:rdopt}
\end{equation}

\noindent where $R(c(\mathbf{x}))$ is the average codeword length. 
Traditionally, lossy compression employs a 2D transform (denoted as $\mathbf{U}$) to produce a new image representation  where image pixels are at least approximately uncorrelated. This process typically generates a vector of transform coefficients as $\boldsymbol{\alpha}=\mathbf{U}^{-1}\mathbf{x}$, such that only few coefficients of $\boldsymbol{\alpha}$ are significantly different from zero. This is critical to achieve good compression performance, and such coefficients can often be interpreted in terms of a frequency representation. 

\subsection{Adaptive transforms for compression}

\begin{figure}
	\centering
	\includegraphics[width=8cm]{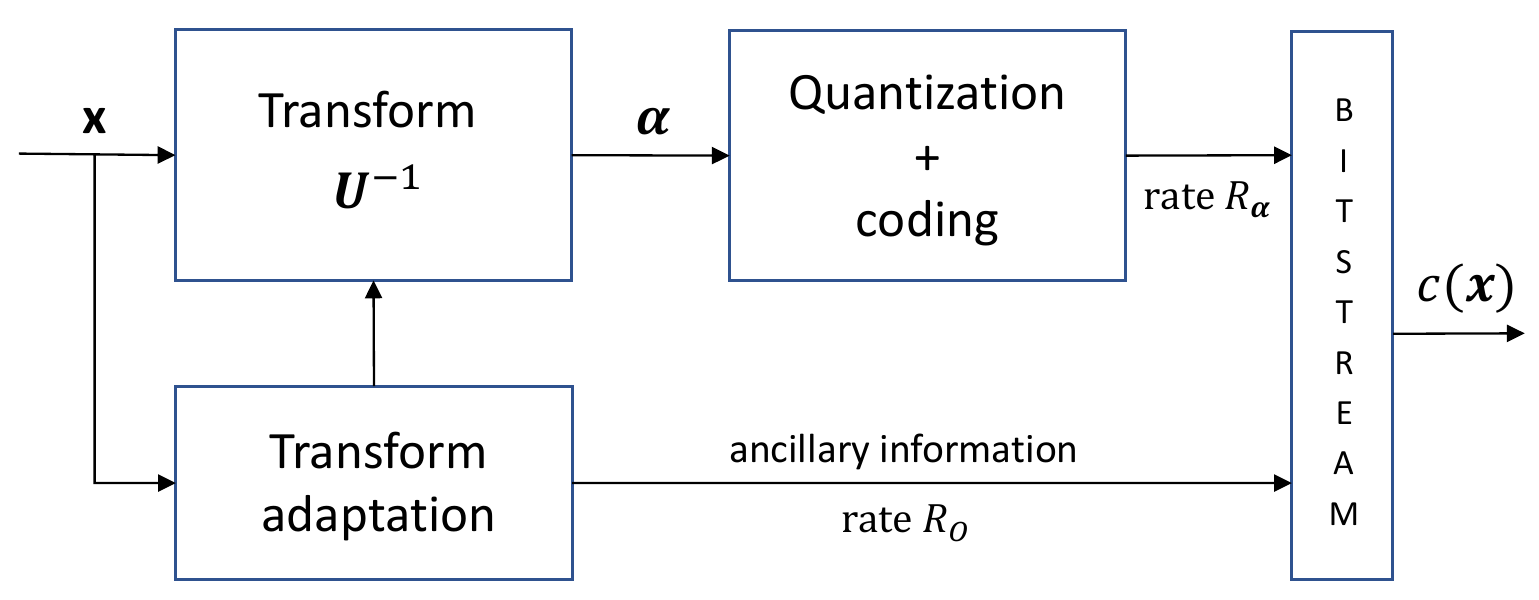} 
	\caption{Block diagram of lossy compression scheme. Coefficients are transformed, quantized and entropy coded. The transform is signal-adaptive, and the bitstream is composed of the coded transform coefficients (rate $R_{\boldsymbol{\alpha}}$) plus the ancillary information (rate $R_O$), for a total rate $R(c(\mathbf{x})) \leq R_T$.}
	\label{fig:lossy}
\end{figure}

More in detail, as  in Fig. \ref{fig:lossy}, the first step consists of the linear transform generating coefficients $\boldsymbol{\alpha}$. Such coefficients are subsequently quantized, and the quantization indexes are losslessly coded using some data compression algorithm such as Huffman  or arithmetic coding. There exist plenty of variations on this scheme, and the interested reader is referred to textbooks on the subject for details, e.g. \cite{sayood2012}. For our discussion, it is important to note that, if the transform to be used is {\em not} known in advance at the encoder and decoder but it is computed adaptively at the encoder in order to optimize the compression process, then some ancillary information has to be communicated to the decoder, in order to reconstruct the corresponding inverse transform to correctly decode the image. The rate term in (\ref{eq:rdopt}) can be written as $R(c(\mathbf{x}))=R_{\boldsymbol{\alpha}}+R_O$, i.e. the rate needed to encode the transform coefficients plus the overhead rate due to the ancillary information; both terms may depend on $\mathbf{x}$, making the design of adaptive transforms a challenging problem.

In this section we focus on the transform stage, as spectral graph theory provides innovative tools to design transforms for image compression. Seeking an ``optimal'' transform has proved to be an elusive goal except for a few rather simple image models. The Karhunen-Lo\`eve transform (KLT) is based on the eigendecomposition of the covariance matrix of the input process and is very similar to the principal component analysis; it has been shown to be optimal for a Gaussian source under mean square error metric and fixed-rate coding \cite{goyal2000}. The DCT \cite{strang1999} is asymptotically equivalent to the KLT for a first-order autoregressive process \cite{jain1979}. However, these models fail to capture the complex and nonstationary behavior typically occurring in digital images, and transform design is still an active research area. While many commonly used transforms, such as the DCT and wavelets \cite{devore1992}, employ a fixed set of basis vectors that need not be communicated to the decoder, the KLT is a signal-adaptive transform. Adaptivity allows to match the basis vectors (the columns of $\mathbf{U}$) to a class of signals of interest, but the transform matrix has to be known at both the encoder and decoder; moreover, the resulting transform has no structure and hence lacks any fast algorithm. These issues have limited the practical use of the KLT for signal compression.

Like the KLT, the GFT is also based on an eigenvector decomposition. The GFT interprets a signal as being defined on a graph, and calculates the eigenvector decomposition of the corresponding graph Laplacian as in (\ref{eq:eigendecomposition}). Thus, while the KLT takes a statistical approach, describing correlations among image pixels through estimates of their linear correlation coefficients, the GFT employs a more flexible approach, in which pixel similarities are encoded into the weights of an undirected graph, where each node of the graph represents a pixel, and each edge weight represents the ``similarity'' of the two pixels at the ends of the edge. The two transforms are related to each other; in particular, \cite{hu15} shows that the GFT approximates the KLT for a piece-wise first-order autoregressive process, while \cite{zhang13} shows that the GFT is optimal for decorrelation of an image following a Gauss-Markov random field (GMRF) model. Both the KLT and GFT can be interpreted in terms of kernels. In the KLT, the covariance matrix (which is PSD) is obtained from a (PSD) linear  kernel, whereas the GFT is obtained from the graph Laplacian matrix, which is also PSD.

In practice, however, a graph can be computed  for each individual image, making the GFT a more flexible framework for transform design. Roughly speaking, the graph in the GFT encodes image structures, as opposed to statistical correlations. This is useful because one can decide the degree of accuracy with which structures are represented in the graph, providing means to reduce the overhead of signaling the transform to the decoder. Referring to Fig. \ref{fig:lossy}, using the GFT in a transform coding scheme requires communication to the decoder a description of the graph as ancillary information; the relatively high overhead requires finding descriptions of the graph that are optimized in a rate-distortion (RD) sense, i.e. they are sufficiently informative to yield effective transforms, without requiring a  large overhead.

\subsection{Graph Fourier Transform and graph design}

\begin{figure}
	\centering
	\includegraphics[width=8cm]{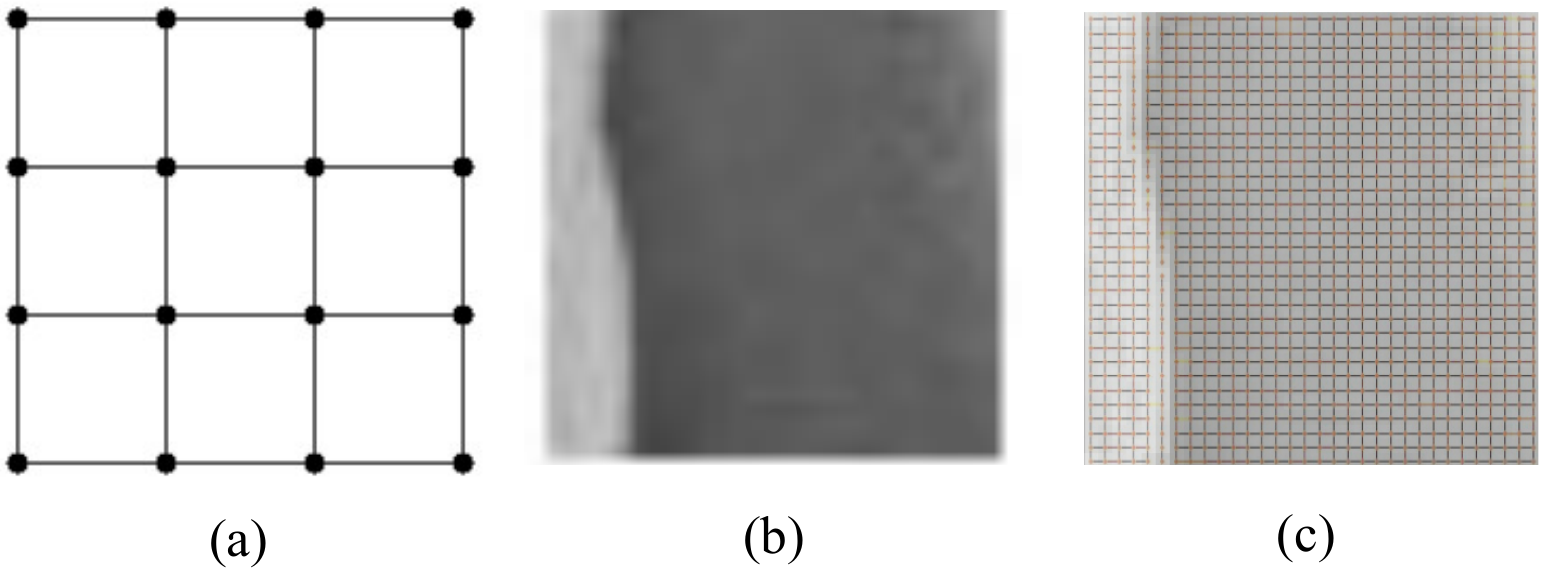}
	\caption{(a) Square grid graph; (b) a 32x32 image block, and (c) an example of graph superimposed onto it. Graph edges with white color denote weak pixel similarity, and the figure shows that graph edges indeed encode structural information about the image.}
	\label{fig:square_grid}
\end{figure}

As has been seen in Sec. \ref{subsec:spectrum}, the graph topology and set of weights $\mathcal{G}(\mathcal{V}, \mathcal{E}, \mathbf{W})$ fully define the graph Laplacian matrix, from which the GFT is computed\footnote{Other approaches are also possible, e.g. \cite{sandryhaila2013} defines a GFT obtained from the eigendecomposition of the graph adjacency matrix.}. Hence, obtaining a ``good'' GFT amounts to selecting the topology and weights yielding the best compression performance in an RD sense as in (\ref{eq:rdopt}). Regarding the topology, given that 2D images are typically defined on a square grid, a square grid graph is typically employed as in Fig. \ref{fig:square_grid}-a, where each pixel is connected to its four horizontal and vertical neighbors. In principle, one may decide to add graph edges corresponding to diagonal neighbors, or connecting pixels whose distance is larger than one. However, this may greatly increase the overhead of communicating the graph, unless edges are carefully selected e.g. as proposed in \cite{rotondo2015}. Fig. \ref{fig:square_grid}-b shows a 32x32 image block with the corresponding graph superimposed onto it, emphasizing the fact that the graph encodes image structures.

\subsubsection{Choosing edge weights}

The weight $w_{i,j}$ on each edge of the graph is conventionally computed as a function of the difference in pixel values $x_i$ and $x_j$ connected by that edge---i.e. the photometric distance---as computed in (\ref{eq:edgeWeight}). 


However, it is easy to realize that real-valued graph weights are too expensive in terms of overhead.
In \cite{shen10pcs,kim2012,fracastoro2015c} the  weights are constrained to be in the set $\{ 1,0 \}$, implying that the graph only describes strong or zero correlation; the weights are chosen from detected image edges \cite{shen2010}, using a greedy optimization algorithm \cite{kim2012}, or from the output of an image segmentation algorithm, where an independent graph with weights equal to one is associated to  each region and the resulting GFT plays the role of a shape-adaptive transform \cite{fracastoro2015c}. In \cite{fracastoro2015d} the difference $|x_i - x_j|$ is quantized to two values using a pdf-optimized uniform quantizer, yielding a graph that is always connected by construction; although weight binarization leads to suboptimal compression efficiency, it is shown that a suitably designed quantizer makes the performance loss very small. In \cite{hu15} two sets of weight values are used, i.e. $w_{i,j} \in \{ 1,0 \}$ for image blocks characterized by strong or zero correlation, and $w_{i,j} \in \{ 1,c \}$ for blocks with strong or {\em weak} correlation. The constant $c$ is optimized using a model suitable for piecewise smooth signals, and very good results are obtained in the compression of depth map images. The overhead incurred by the graph, however, makes it harder to obtain significant gains on natural images. This problem is addressed in \cite{fracastoro2015d}, where edge prediction followed by coding is used to reduce the overhead, leading to performance gains between 1 and 3 dB in peak signal-to-noise ratio (PSNR) over the DCT. More sophisticated graph coding techniques may also be devised, e.g. one might in principle apply contour coding techniques as in \cite{zheng2017,verdoja2017} to reduce the cost of representing the graph. Moreover, in \cite{verdoja2017} directional graph weight prediction modes are proposed, which avoid transmitting any overhead information to the decoder.

\subsubsection{Graph learning}

Defining a good graph from data observations is so important in many applications, and particularly in compression, that more structured methods have been developed to this purpose; this problem is referred to as {\em graph learning}. In \cite{pavez2016}, the authors formulate the graph learning problem as a precision matrix estimation with generalized Laplacian constraints. In \cite{dong2016}, a sparse combinatorial Laplacian matrix is estimated from the data samples under a smoothness prior. In \cite{pavez2015}, a new class of transforms called graph template transform is proposed; the authors use a graph template to impose a sparsity pattern and approximate the empirical inverse covariance based on that template. 

While the methods above are effective at deriving a graph from data, none of them takes into account the actual cost of representing, and thus coding, the graph, which is clearly a major problem for image compression. In \cite{fracastoro2016b} a novel graph-based framework is proposed,  explicitly accounting for the cost of transmitting the graph. The authors treat the edge weights $w_{i,j}$ as a graph signal that lies on the dual graph. They compute the GFT of the weights graph and code its quantized transform coefficients. The choice of the graph is posed as a RD optimization problem.

\subsubsection{Reducing GFT complexity}

Besides the cost required to represent and encode the graph, the complexity of solving (\ref{eq:eigendecomposition}) to obtain the GFT matrix may outweigh any obtained coding gain. Indeed, applying the GFT to large blocks may quickly become infeasible. In \cite{hu15} the authors propose to use a lookup table storing the GFTs for the most commonly used graphs, so that only the index of the corresponding chosen transform has to be transmitted; this has been shown to work well for relatively small block sizes. Moreover, in \cite{hu12icip,hu15} it is proposed to apply the GFT to a low-resolution version of the image, and to employ edge-adaptive filtering to restore the original resolution. In \cite{egilmez2016} graph-based separable transforms are proposed, where the transform is optimized separately along rows and columns. In \cite{lu2016} symmetric line-graph transforms are proposed, in which symmetries are exploited to reduce the number of operations needed to compute the transform.

\subsubsection{Compression performance}

\begin{figure}
	\centering
	\includegraphics[width=3.5cm]{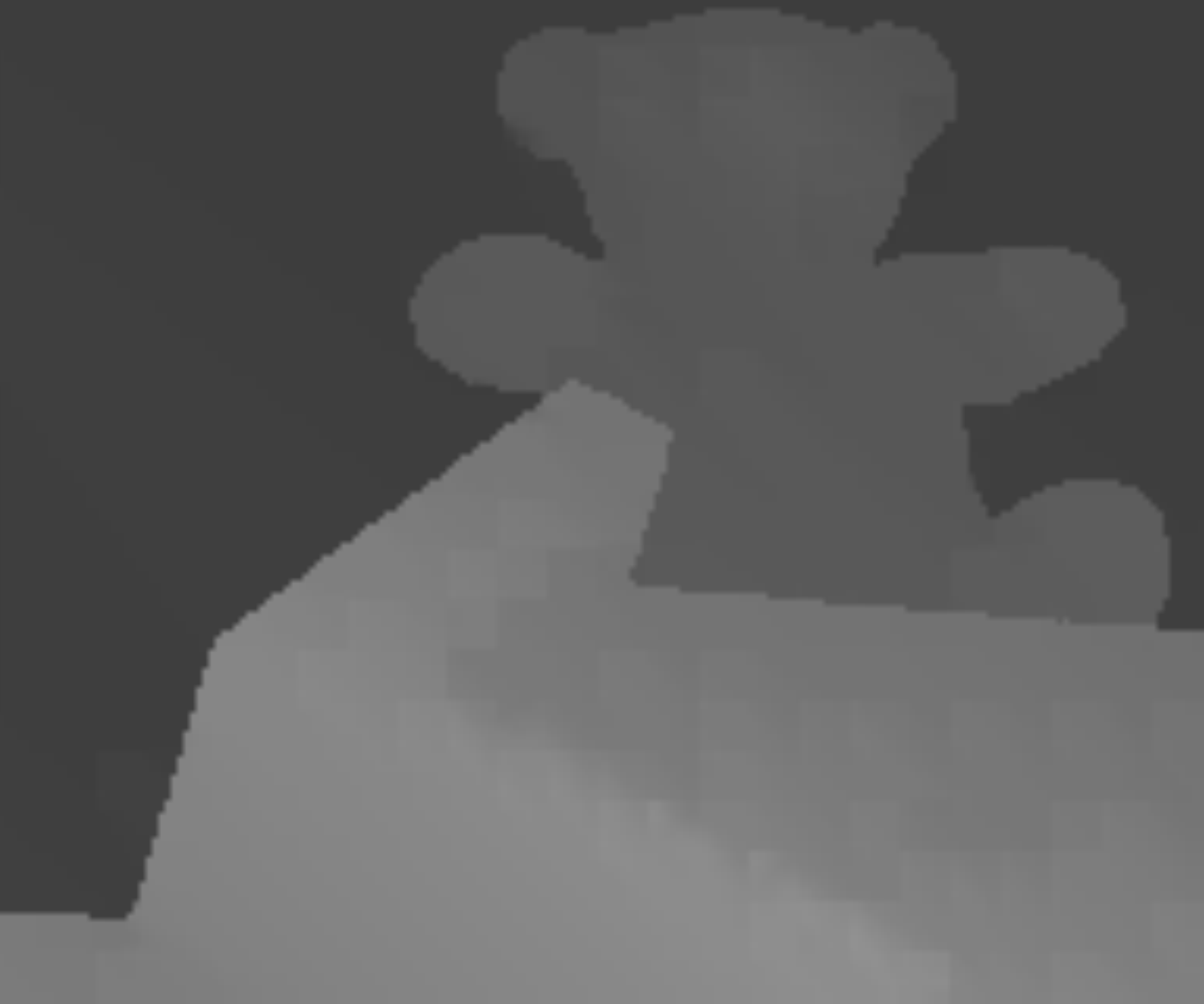} 
	\includegraphics[width=3.5cm]{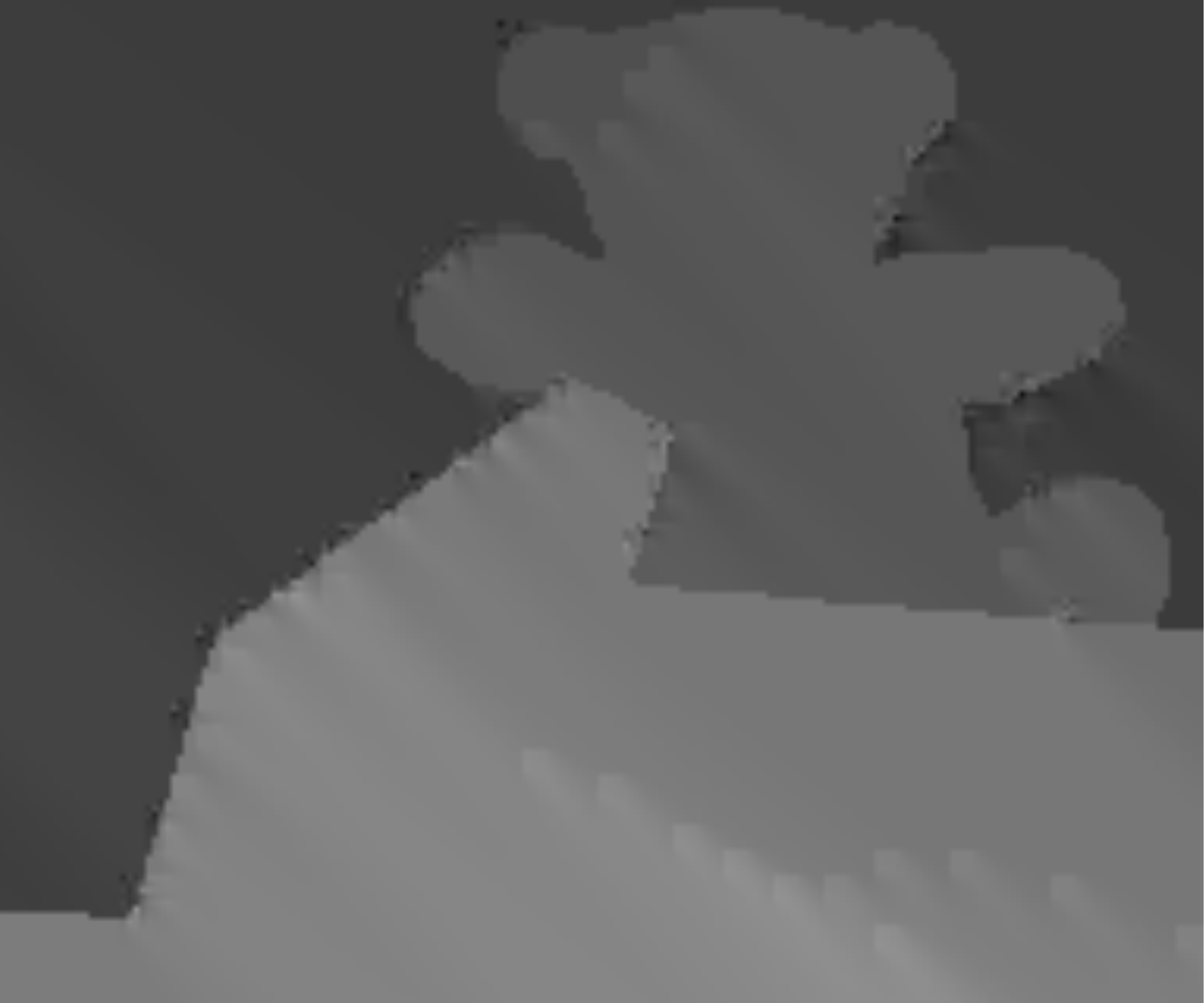} \\ \vspace{0.2cm}
	\includegraphics[width=6cm]{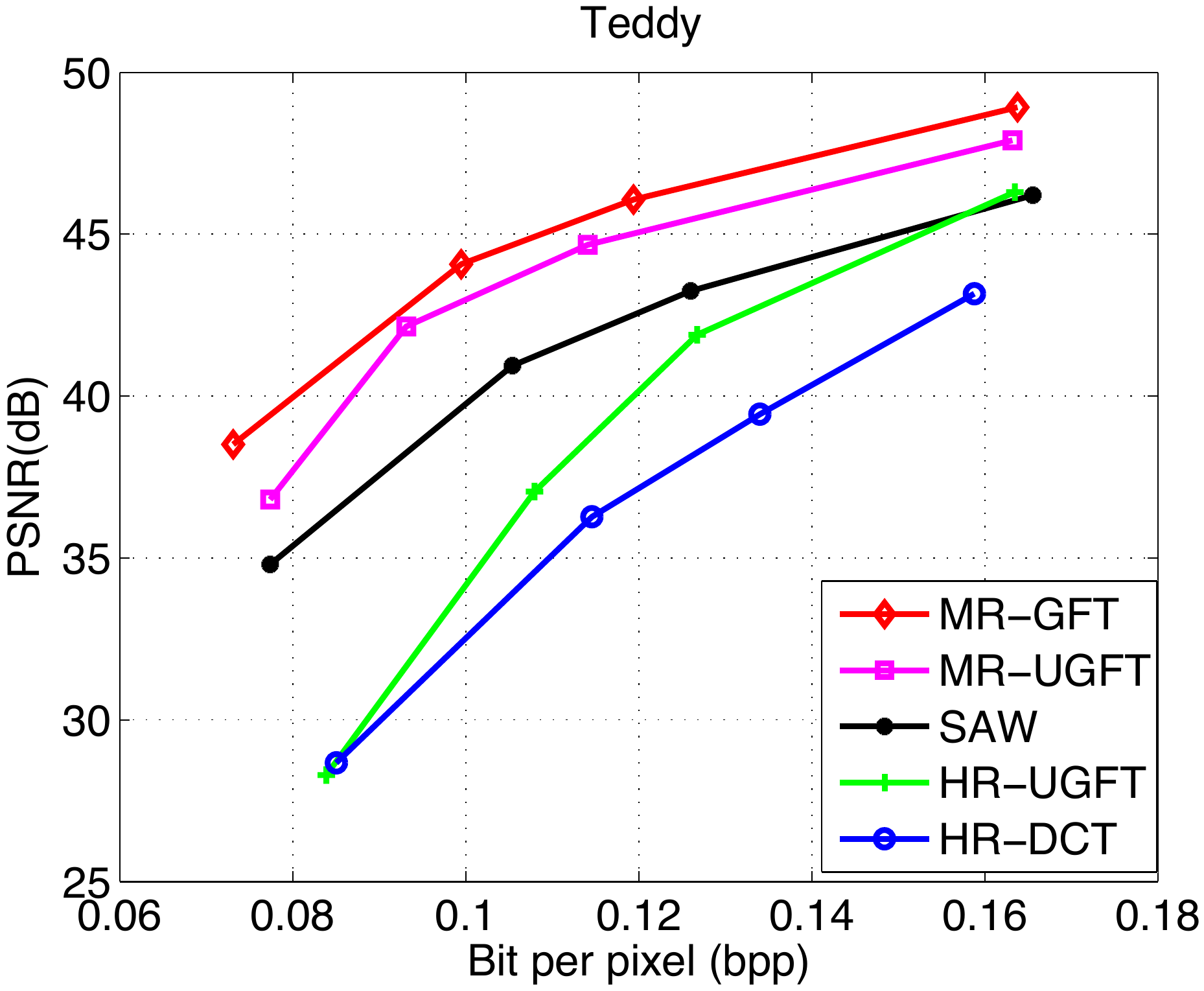}
	\caption{Compression performance on the {\em Teddy} depth image, employing the MR-GFT \cite{hu15}, the MR-UGFT \cite{hu12icip}, H.264/AVC in intra mode (HR-DCT), the HR-UGFT \cite{shen10pcs}, and the shape-adaptive wavelet (SAW, \cite{maitre2010}). The top line shows reconstruction using the MR-GFT (left) and HR-DCT (right) at 0.1 bpp.}
	\label{fig:rd_curves}
\end{figure}

Several authors have applied the GFT for image and video compression. In a practical setting, GFT coefficients over different image blocks may correspond to different frequencies (eigenvalues), making entropy coding somewhat more difficult. A possible solution employs bit-plane coding of coefficient significance, which depends only on the energy distribution of the transform coefficients. In Fig. \ref{fig:rd_curves} we report a comparison on compression of depth images. As can be seen, the MR-GFT codec \cite{hu15} outperforms the other transforms in rate-distortion sense; in particular, gains between 5 and 10 dB are obtained with respect to the corresponding DCT-based coder. Correspondingly, at the same bit-rate the MR-GFT yields a depth image with less evident artifacts. 


\subsection{Steerable transforms from GFT}

\begin{figure}[t]
	\centering
	\includegraphics[width=8cm]{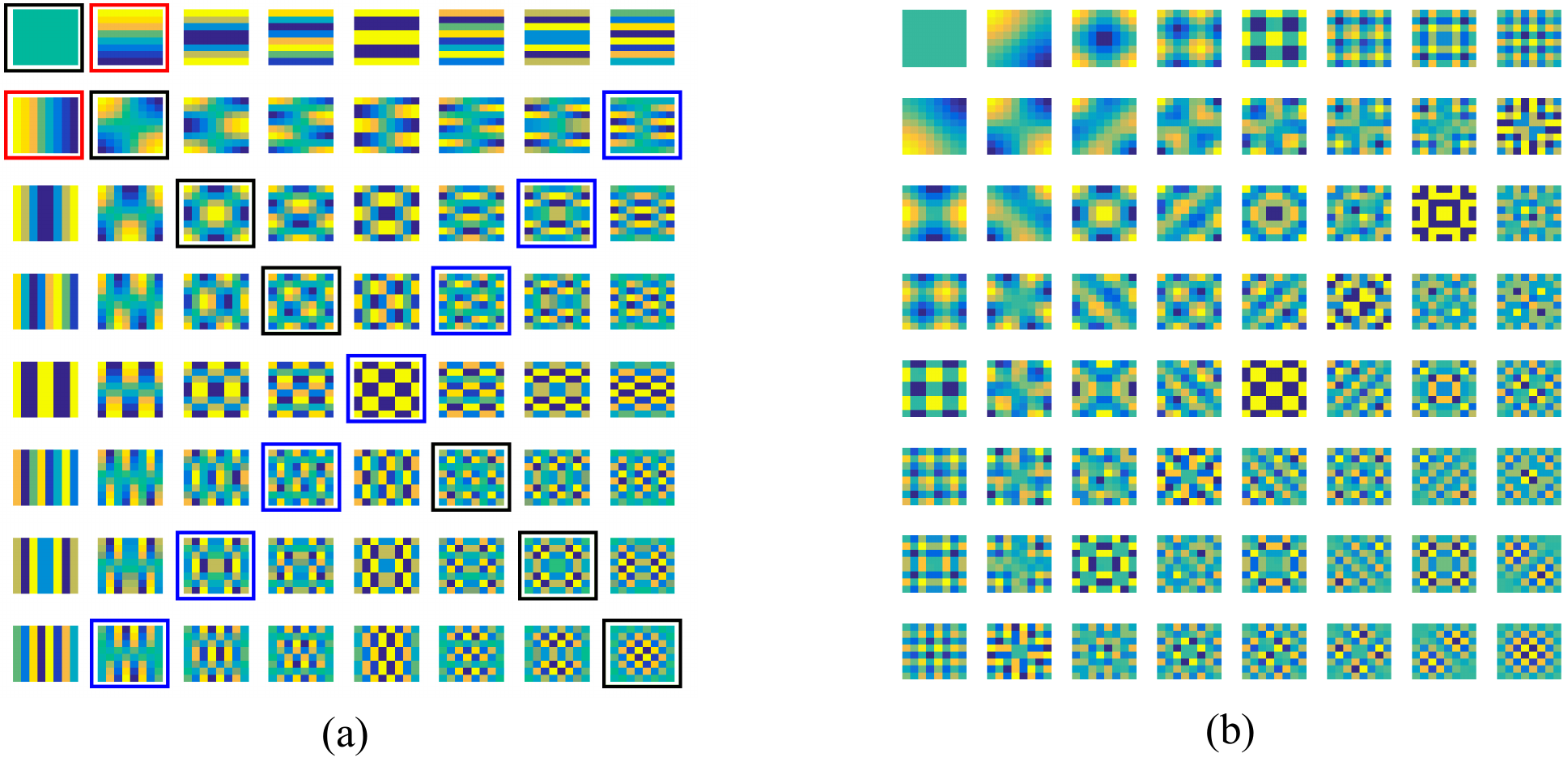}
	\caption{(a) 2D-DCT basis vectors represented in matrix form (with $n=8$): the corresponding two eigenvectors of an eigenvalue with multiplicity 2 are highlighted in red, the $n-1$ eigenvectors corresponding to $\lambda=4$ are highlighted in blue and the $n-1$ eigenvectors corresponding to the eigenvalues with algebraic multiplicity 1 are highlighted in green. (b) Basis vectors of steered 2D-DCT with $\theta_{k,l}=\frac{\pi}{4} \; \forall k,l$.}
	\label{fig:basisDCT}
\end{figure}

In many cases much of the content of an image block can be described by few main structures, employing a simplified image model with much fewer parameters, leading to reduced overhead. In particular, the directional model has become rather popular, e.g. in directional intra prediction modes \cite{wiegand2003} and directional transforms \cite{xu2010,xu2007,zeng2008,chang2008,kamisli2009,cohen2010,dremeau2010}, including more sophisticated transforms such as bandelets \cite{lepennec2005} and anisotropic transforms \cite{peng2010}. The GFT framework can also be employed to design simplified adaptive transforms. As has been seen in Sec. \ref{sec:prelim}, the DCT is the GFT of the line graph with all weights equal to 1.
In the same way, the basis vectors of the 2D-DCT are eigenvectors of the Laplacian of a square grid graph as in Fig. \ref{fig:square_grid} \cite{zhang13}, but the solution to (\ref{eq:eigendecomposition}) for a square grid graph is not unique because the eigenvalues of $\mathbf{L}$ do not all have algebraic multiplicity equal to one \cite{fracastoro2015}. Using the pair $(k,l)$ with $k,l \in [1,n]$ instead of index $i$ in order to emphasize the bidimensionality of the basis vectors corresponding to the eigenvectors of $\mathbf{L}$ in the 2D case, it is easy to show that $\lambda_{k,l}=\lambda_{l,k}$ for $k\neq l$, i.e. these eigenvalues have multiplicity 2. Moreover, $\lambda_{k,n-k}=4$ for $1\le k\le n-1$, i.e. this eigenvalue has multiplicity $n-1$. Graphically this is shown in Fig. \ref{fig:basisDCT}(a), where the basis vectors highlighted in red represent an example of eigenvectors corresponding to the same eigenvalue.

Therefore, the set of all possible eigenbases satisfying (\ref{eq:eigendecomposition}) for the Laplacian of a square grid graph can be represented as 
\begin{equation}
\begin{bmatrix}
\mathbf{u}'_{(k,l)}\\
\mathbf{u}'_{(l,k)}
\end{bmatrix}
=
\begin{bmatrix}
\cos\theta_{k,l} & \sin\theta_{k,l}\\
-\sin\theta_{k,l} & \cos\theta_{k,l}
\end{bmatrix}
\begin{bmatrix}
\mathbf{u}_{(k,l)}\\
\mathbf{u}_{(l,k)}
\end{bmatrix},
\label{eq:rot}
\end{equation}
\noindent where $\mathbf{u}_{(k,l)}$ are the eigenvectors corresponding to the basis vectors of the separable 2D-DCT. Indeed, (\ref{eq:rot}) applies a rotation of an arbitrary angle $\theta_{k,l}$ to each pair of basis vectors  $\mathbf{u}_{(k,l)}$ and  $\mathbf{u}_{(l,k)}$. The new transform is defined by the new eigenvectors $\mathbf{u}'_{(k,l)}$, or equivalently and more handily by the original eigenvectors $\mathbf{u}_{(k,l)}$ plus the set of rotation angles $\theta_{k,l}$. 
Fig. \ref{fig:basisDCT}(b) shows the resulting set of 2D basis vectors when $\theta_{k,l}=\frac{\pi}{4} \; \forall k,l$.

Such angles must be chosen to match the directional characteristics of the image block the transform is applied to, and to minimize the overhead of transmitting the angles. In \cite{fracastoro2015} the same angle is chosen for all pairs of basis vectors with multiplicity equal to 2, i.e. $\theta_{k,l}=\theta$. 
In \cite{fracastoro2016b} $\theta_{k,l}$ is chosen individually for each pair, almost halving the number of DCT coefficients to be transmitted. In \cite{fracastoro2015b} the angles are chosen in a RD optimized fashion. In terms of implementation, in \cite{fracastoro2015b} it is noted that the coefficients of the steered transform can be obtained from the coefficients of the separable 2D-DCT of the image block, followed by the application of a sparse rotation matrix; this makes the complexity only marginally higher than that of the separable 2D-DCT. Interestingly, the same principle can be applied  to other transforms as well. In \cite{fracastoro2017} it is shown that steerable 1D and 2D Discrete Fourier transforms (DFT) can be obtained. In the one-dimensional case, 
rotations change the balance of signal energy between the real and imaginary parts of the DFT; the resulting transform is related to the DCT, the discrete sine transform and the Hilbert transform. In 2D, 
rotations indeed correspond to geometric rotations.

\subsection{Applications}

We have previously mentioned applications of GFT to the compression of depth maps and natural images. Other authors have applied various types of GFT to video compression. In \cite{hu15spl} the GFT is optimized for intra-prediction residues, while in \cite{chao2016} the authors propose a block-based lifting transform on graphs for intra-predicted video coding. A graph-based method for inter-predicted video coding has been introduced in \cite{egilmez2015}, where the authors design a set of simplified graph templates capturing basic statistical characteristics of inter-predicted residual blocks. In \cite{lu2016} symmetric line-graph transforms are proposed for predictive video coding. In \cite{chao2016b} a new edge coding methods is introduced with application to intra prediction residuals.

Applications of GFT to other types of data have also been presented. 
In \cite{motz2016} a graph-based representation is applied to the problem of interactive multiview streaming, while in \cite{su2017} a weighted GFT is employed for compression of light fields. In \cite{thanou2017} the time-varying geometry of 3D point cloud sequences is represented as a set of graphs on which motion estimation is performed, whereas in \cite{maugey2014} the graph representation is used to encode luminance information in multiview video.

Finally, a few works have employed graph wavelets to image/video coding problems. In \cite{narang2013b} a graph wavelet transform has been proposed for image compression. In \cite{martinez2011,martinez2011b,martinez2016} the authors propose a complete video encoder based on lifting-based wavelet transforms on graphs; constructing a graph in which any pixel could be linked to several spatial and temporal neighbors, they jointly exploit spatial and temporal correlation. In \cite{chao2015} lifting-based graph wavelets are applied to compression of depth maps. In \cite{zeng2017} graph wavelets are employed for the compression of hyperspectral images. These 3D images are characterized by a significant amount of correlation among images at different wavelengths, as well as spatial correlation, which are exploited constructing a spatial-spectral graph for groups of bands.

\section{Graph-based Image Restoration}
\label{sec:restore}
Image restoration is an inverse problem; given a noise-corrupted and/or degraded observation $\mathbf{y}$, one is tasked with restoring the original signal $\mathbf{x}$. 
Examples of restoration problems include image denoising, interpolation, super-resolution, deblurring, etc. 
An example generic image formation model is:
\begin{equation}
\mathbf{y} = \mathbf{H} \mathbf{x} + \mathbf{z}
\label{eq:restoreModel}
\end{equation}
where $\mathbf{H}$ is a degradation matrix that performs down-sampling, blurring etc., and $\mathbf{z}$ is an additive noise. 

Image restoration is an ill-posed problem, and thus prior knowledge about the sought signal $\mathbf{x}$ is required to regularize the problem. 
In this section, we describe recent graph-signal priors and their usages in the literature for image restoration. 

\subsection{Image Denoising}
\label{subsec:denoise}

We start with image denoising, which is the most basic image restoration problem with image formation model (\ref{eq:restoreModel}) where $\mathbf{H} = \mathbf{I}$, and $\mathbf{z}$ is typically assumed to be an \textit{additive white Gaussian noise} (AWGN). 

Using a Bayesian approach, a typical \textit{maximum a posteriori} (MAP) formulation has the following form:
\begin{equation}
\min_{\mathbf{x}} 
\| \mathbf{y} - \mathbf{x} \|^2_2 + \mu \, R(\mathbf{x})
\label{eq:denoising}
\end{equation}
where $R(\mathbf{x})$ is the negative log of a signal prior or regularization term for candidate signal $\mathbf{x}$, and $\mu$ is a weight parameter. 
The crux is to define a prior $R(\mathbf{x})$ that discriminates target signal $\mathbf{x}$ against other candidates, while keeping optimization (\ref{eq:denoising}) computationally efficient. 
There have been many priors $R(\mathbf{x})$ proposed with a varying degree of success; \textit{e.g.} \textit{total variation} (TV) \cite{rudin92}, kernel regression \cite{takeda07}, nonlocal means (NLM) \cite{buades05}, sparsity with respect to a pre-defined over-complete dictionary \cite{elad06}, etc. 
We discuss popular graph-signal priors in the literature, where the underlying graphs are often signal-adaptive.

Note that one may choose not to pose a MAP optimization like (\ref{eq:denoising}) at all; \cite{milanfar13} argued it is more direct to address image denoising as a filtering problem: 
\begin{equation}
\mathbf{x} = \mathbf{D}^{-1} \mathbf{W} \mathbf{y}
\label{eq:noiseFilter}
\end{equation}
where $\mathbf{D}^{-1} \mathbf{W}$ is row-stochastic, and filter coefficients in $\mathbf{W}$ are designed adaptively based on local / non-local statistics\footnote{Instead of explicitly normalization, a recent work \cite{milanfar16} shows that an image filter can be approximately normalized with lower complexity.}. 
While graph-based filters derived from (\ref{eq:denoising}) can often be casted in the same framework in \cite{milanfar13}, we instead focus on the introduction of graph-based priors $R(\mathbf{x})$ for (\ref{eq:denoising}). 
We refer interested readers in image denoising using (\ref{eq:noiseFilter}) to the extensive overview paper \cite{milanfar13}.

\subsubsection{Sparsity of GFT Coefficients}

One conventional approach is to map an observed signal $\mathbf{y}$ to a pre-selected transform domain, and assuming sparse signal representation in the domain, perform hard / soft thresholding on the transform coefficients  \cite{donoho07}.
Instead of pre-determined transforms and wavelets, one can use graph transforms and wavelets as basis and perform coefficient thresholding subsequently. 
Probabilistically, \cite{zhang15} showed that the graph Laplacian can be roughly interpreted as an inverse covariance matrix of a \textit{Gaussian Markov Random Field} (GMRF), and thus the corresponding GFT is equivalent to the \textit{Karhunen Lo\`{e}ve Transform} (KLT) that deccorelates an input random signal.  
Hence it is reasonable to assume that an appropriately chosen GFT can sparsify a signal, resulting in a smaller $l_0$-norm. 

As a concrete implementation, \textit{non-local graph based transform} (NLGBT) \cite{hu13} used GFT for depth image denoising as follows.
Assuming self-similarity in images as done in NLM \cite{buades05} and BM3D \cite{dabov07}, $N-1$ similar patches $\mathbf{y}_i$, $i \geq 2$, to a target patch $\mathbf{y}_1$ are first searched in the depth image, in order to compute an average patch $\bar{\mathbf{y}}$. 
Assuming a four-connected graph that connects each pixel to its four nearest neighbors, the weight $w_{i,j}$ of an edge connecting pixels $i$ and $j$ is computed using (\ref{eq:edgeWeight}). 
Note that the edge weights are computed using photometric distance, making the resulting filter signal-adaptive, thus improving its performance \cite{milanfar13}.

It is legitimate to ask how sensitive would the computed edge weights in (\ref{eq:edgeWeight}) are to noise in observations. 
If one uses a pre-filtered version of the observation to compute edge weights using (\ref{eq:edgeWeight}) \cite{milanfar13}, then it is shown that the computed eigenvectors are robust to noise \cite{meyer14}.
In \cite{liu15icassp}, the authors performed low-pass filtering on computed edge weights in a dual graph as pre-filtering.  
In \cite{hu16spl}, for piecewise smooth (PWS) images, the authors minimize the total variation of edge weights in a dual graph.
In \cite{hu13}, the averaging over $N$ patches effectively constitutes one low-pass pre-filtering. 

Given graph Laplacian $\mathbf{L}$ computed for the constructed graph, GFT $\mathbf{U}^T$ is computed as the basis that spans the signal space. 
The $N$ similar patches are denoised jointly as follows:
\begin{equation}
\min_{\boldsymbol{\alpha}} 
\sum_{i=1}^N \|\mathbf{y}_i - \mathbf{U} \boldsymbol{\alpha}_i \|_2^2 + 
\tau \sum_{i=1}^N \| \boldsymbol{\alpha}_i \|_0
\label{eq:weiDenoise}
\end{equation}
where the weight parameter $\tau$ can be estimated using \textit{Stein Unbiased Risk Estimator} (SURE) 
\cite{luisier07}.
Soft thresholding is used to iteratively minimize the second term. 
Shrinkage of transform coefficients for image denoising is common \cite{iizuka14}.
If the $l_0$-norm is replaced by a convex $l_1$-norm, then fast algorithms such as the split Bregman method \cite{goldstein09} can be used.

(\ref{eq:weiDenoise}) is solved iteratively, where between iterations edge weights are updated in (\ref{eq:edgeWeight}) using computed solution in (\ref{eq:weiDenoise}). 
\cite{hu13} showed that for PWS images, the performance can out-perform state-of-the-art algorithms like BM3D \cite{dabov07}.

\subsubsection{Graph Laplacian regularizer}

Another common graph-signal prior is the \textit{graph Laplacian regularizer} $R(\mathbf{x}) = \mathbf{x}^T \mathbf{L} \mathbf{x}$; it can be interpreted as a \textit{Tikhonov regularizer} $\| \Gamma \mathbf{x} \|_2^2$ where $\Gamma = \mathbf{U} \mathbf{\Lambda}^{1/2} \mathbf{U}^T$ given $\mathbf{L} = \mathbf{U} \mathbf{\Lambda} \mathbf{U}^T$.
From (\ref{eq:smoothness}), minimizing $\mathbf{x}^T \mathbf{L} \mathbf{x}$ means that connected pixel pairs $(i,j)$ by large edge weights $w_{i,j}$ will have similar sample values, or that the energy of the signal resides mostly in the low frequencies. 
$\mathbf{x}^T \mathbf{L} \mathbf{x}$ for restoration is prevalent across many fields, such as graph-based classifier in machine learning \cite{belkin04}.  


Using $R(\mathbf{x}) = \mathbf{x}^T \mathbf{L} \mathbf{x}$ in (\ref{eq:denoising}) leads to the following optimal solution $\mathbf{x}^*$: 
\begin{equation}
\mathbf{x}^* = 
\mathbf{U} \mathrm{diag} 
\left( \frac{1}{(1+ \mu \lambda_1)}, \ldots, \frac{1}{(1+ \mu \lambda_N)} \right) 
\mathbf{U}^T \mathbf{y}
\label{eq:MAPsoln}
\end{equation}
The resulting low-pass filter on $\mathbf{y}$ in GFT domain---smaller filter coefficient $(1+\mu \lambda_i)^{-1}$ for larger $\lambda_i$---can be implemented efficiently using Chebychev polynomial approximation, as discussed in Section\;\ref{sec:filter}. 

Alternatively, \cite{chen14} defined signal smoothness using (\ref{eq:smoothness3}), and, assuming that the Hermetian of the weight matrix $\mathbf{\mathbf{W}}^* = h(\mathbf{W})$ is a polynomial of $\mathbf{W}$, then the optimal MAP denoising filter with a $l_2$-norm fidelity term is derived as $g(\lambda_n)$ without matrix inversion:
\begin{equation} 
g(\lambda_n) = \frac{1}{1 + \mu (1 - \lambda_n)^2}
\end{equation}
See \cite{chen14} for details. 


It is known that the graph Laplacian can be derived from sample points of a differentiable manifold, and if the samples are randomly distributed, then the graph Laplacian operator converges to the Laplace-Beltrami operator in continuous manifold space when the number of samples tends to infinity \cite{belkin05}. 
The graph Laplacian regularizer can also be interpreted from a continuous manifold perspective \cite{pang17}, with additional insights that connect the prior to TV. 
Because edge weights $w_{i,j}$ in (\ref{eq:edgeWeight}) are typically defined signal-adaptively, it is appropriate to write the prior as $\mathbf{x}^T \mathbf{L}(\mathbf{x}) \mathbf{x}$.  
More generally, $w_{i,j}$ can be computed as the Gaussian of the difference in a set of pre-defined \textit{exemplar functions} $f(\,)$ evaluated at node $i$ and $j$.
Examples of $f(\,)$ can be the $x$- and $y$-coordinates of a pixel, and intensity value of the pixel.

\begin{figure}[!t]
\centering
{\includegraphics[width=80pt]{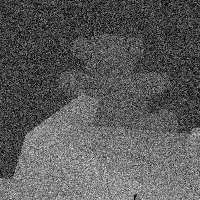}}\hspace{2pt}
{\includegraphics[width=80pt]{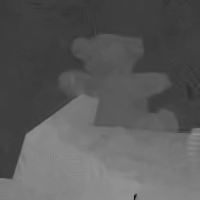}}\hspace{2pt}
{\includegraphics[width=80pt]{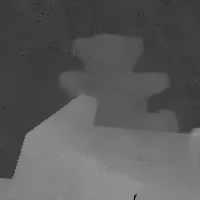}}\\\vspace{2pt}
\stackunder[5pt]{\includegraphics[width=80pt]{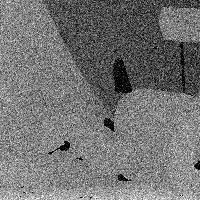}}{Noisy, 18.60\,dB}\hspace{2pt}
\stackunder[5pt]{\includegraphics[width=80pt]{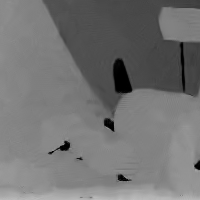}}{BM3D, 33.20\,dB}\hspace{2pt}
\stackunder[5pt]{\includegraphics[width=80pt]{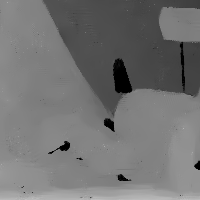}}{OGLR, 34.55\,dB}
\caption{Denoising of the depth image {\it Teddy}, where the original image is corrupted by AWGN with $\sigma_{\cal I}=30$. Two cropped fragments of each image are presented for comparison.}
\label{fig:result_teddy}
\end{figure}

If we view the graph-signal as samples on a continuous manifold, then as the number of samples tends to infinity and the distances among neighboring samples go to $0$, $\mathbf{x}^T \mathbf{L}(\mathbf{x}) \mathbf{x}$ converges to a continuous functional \cite{pang17},
\begin{equation}
\int_{\Omega} 
\nabla x^T \mathbf{G}^{-1} \nabla x 
\left( \sqrt{\mathrm{det}(\mathbf{G})}\right)^{2 \gamma - 1}
d \mathbf{s} 
\label{eq:contFunc}
\end{equation}
where $\mathbf{G}$ is defined as follows:
\begin{equation}
\mathbf{G} = \sum_{n=1}^n \nabla f_n \nabla f_n^T
\end{equation}
$\mathbf{G}$ can be viewed as the \textit{structure tensor} of the gradient of the exemplar functions $\left\{ \nabla f_n \right\}_{n=1}^N$. 
For convenience, define now $\mathbf{D} = \mathbf{G}^{-1} \left( \sqrt{\mathrm{det}(\mathbf{G})}\right)^{2 \gamma - 1}$. 
\cite{pang17} then showed that the solution to the continuous counterpart of optimization (\ref{eq:denoising}) can be implemented as an anisotropic diffusion:
\begin{equation}
\partial_t x = \mathrm{div} \left( \mathbf{D} \nabla x^* \right) 
\end{equation}
$\mathbf{D}$ in this context is also the \textit{diffusivity} that determines how fast an image is being diffused. 
For $\gamma < 1$, through eigen-analysis of $\mathbf{D}$ one can show that the diffusion process is divided into two steps: i) a forward diffusion process that \textit{smooths} along an image edge, and ii) a backward diffusion process that \textit{sharpens} perpendicular to an image edge. 
When $\gamma = 1$, the diffusion process is analogous to TV in the continuous domain. 
This explains why denoising using the graph Laplacian regularizer $\mathbf{x}^T \mathbf{L}(\mathbf{x}) \mathbf{x}$ works particularly well for PWS images, such as depth images shown in Fig.\;\ref{fig:result_teddy}. 


Orthogonally, \cite{shi16} proposed a fast graph Laplacian implementation of \textit{low dimensional manifold model} (LDMM) \cite{osher16}.
In particular, LDMM \cite{osher16} assumes that the size-$d$ pixel patches of an image are points in a $d$-dimensional space that lie in a low-dimensional manifold, commonly called \textit{patch manifold}.
Thus the dimensionality of the manifold can be used as a prior to regularize an inverse problem:
\begin{equation}
\min_{\mathbf{x} \in \mathbb{R}^{m \times n}, \mathcal{M} \subset \mathbb{R}^d}
\mathrm{dim}(\mathcal{M})
~~~ \mbox{s.t.} ~ 
\mathbf{y} = \mathbf{x} + \mathbf{z}, 
~~ P(\mathbf{x}) \subset \mathcal{M}
\label{eq:LDMM}
\end{equation}
where $\mathbf{y}$, $\mathbf{x}$ and $\mathbf{z}$ are the observed image, target image and noise respectively, $P(\mathbf{x})$ are the patches of image $\mathbf{x}$, and $\mathcal{M}$ is the patch manifold. 
\cite{osher16} showed that the dimensionality of the manifold can be written as a sum of \textit{coordinate functions}. For any $\mathbf{x} \in \mathcal{M}$: 
\begin{equation}
\mathrm{dim}(\mathcal{M}) = 
\sum_{j=1}^d \|  \nabla_{\mathcal{M}}  \alpha_j(\mathbf{x}) \|^2
\end{equation}
where $\alpha_i(\mathbf{x})$ is the $i$-th coordinate function.
(\ref{eq:LDMM}) can be solved iteratively, where the key step to solve for the new image $\mathbf{x}^{k+1}$ and coordinate functions $\alpha_i^{k+1}$ requires a point integral method \cite{li16} that is computationally complex.
Instead, \cite{shi16} proposed to use a \textit{weighted graph Laplacian} (WGL) method to replace the point integral method:
\begin{align}
\min_u & \sum_{\mathbf{x} \in P \setminus S} 
\left( \sum_{\mathbf{y} \in P} w(\mathbf{x}, \mathbf{y})(u(\mathbf{x}) - u (\mathbf{y}))^2 \right)
\nonumber \\
& + \frac{|p|}{|S|} \sum_{\mathbf{x} \in S}
\left( \sum_{\mathbf{y} \in P} w(\mathbf{x},\mathbf{y})(u(\mathbf{x}) - u(\mathbf{y}))^2 \right)
\end{align}
The corresponding Euler-Lagrange equation is a linear system that is symmetric and positive definite.
This is much easier to solve than the point integral method.

\subsubsection{Graph Total Variation}

Instead of the graph total variation (\ref{eq:smoothness2}) defined in \cite{chen15}, there exist works \cite{elmoataz08,hidane13,couprie13,berger17} that defined and optimized TV for graphs in a more traditional manner as the seminal work \cite{rudin92}\footnote{\cite{couprie13} actually defines a more general notion called \textit{dual constrained total variation} (DCTV) that includes TV as a special case, and proposed a parallel proximal algorithm as solution.}.
Specifically, \textit{local gradient} $\nabla_i \mathbf{x} \in \mathbb{R}^N$ at a node $i \in \mathcal{N}$ is first defined:
\begin{equation}
\left(\nabla_i \x \right)_j = (x_j - x_i) W_{i,j}
\end{equation}
Then the (isotropic) graph total variation is defined as follows:
\begin{equation}
\| \mathbf{x} \|_{\mathrm{TV}} = \sum_{i \in \mathcal{V}} \| \nabla_i \x \|_2
= \sum_{i \in \mathcal{V}} \sqrt{ \sum_{j \in \mathcal{V}} (x_j - x_i)^2 W_{i,j}^2}
\label{eq:TVNorm}
\end{equation}
Because the TV-norm is convex but non-smooth, there exist specialized algorithms that minimize it with a fidelity term, such as proximal gradient algorithms \cite{hidane13,couprie13}.

As an illustrative example, in \cite{berger17} a signal reconstruction given noise samples $\y$ from sampling matrix $\mathbf{S}$ is formulated as:
\begin{equation}
\min_{\x \in \mathbb{R}^N} \| \x \|_{\mathrm{TV}} ~~~
\mbox{s.t.} ~~ \|\y - \mathbf{S} \x \|_2 \leq \epsilon
\label{eq:tvDenoise}
\end{equation}
To solve (\ref{eq:tvDenoise}), the authors first convert the $L_1$-norm to its convex conjugate---a $L_{\infty}$-norm ball---leading to a saddle point formulation, similarly done in \cite{hidane13}.
Then they use a first-order primal-dual algorithm \cite{chambolle11}, since the new formulation has proximal operators that are much easier to compute.
A distributed version of the algorithm is also provided when handling a large graph.
Experimental results show that optimization of this graph TV norm (\ref{eq:TVNorm}) has better performance that earlier defined smoothness notions (\ref{eq:smoothness}) and (\ref{eq:smoothness2}).
See \cite{berger17} for details.

\subsubsection{Wiener Filter}

More recently, instead of relying on a MAP formulation with sparsity or smoothness priors for regularization, one can approach the denoising problem from a statistical point of view and design a \textit{Wiener filter} that minimizes the mean square error (MSE) instead \cite{yagan16,perraudin17}.
In particular, \cite{perraudin17} first generalizes the notion of \textit{wide-sense stationarity} (WSS) for graph-signals (with generalized translation and modulation operators on graphs \cite{shuman16vf}), estimates the \textit{power spectral density} (PSD), and computes the minimal MSE (MMSE) graph Wiener filter. 
There are several advantages to employ a Wiener filter approach. 
First, unlike the smoothness prior that assumes implicitly a GMRF signal model, as long as the PSD can be robustly estimated, the Wiener filtering approach is more general and does not require a Gaussian assumption.
Second, there is no need to tune a weight parameter ($\mu$ in (\ref{eq:denoising})) to trade off the fidelity term with the prior term.
Third, the specificity of the estimated PSD per graph frequency can be exploited during denoising. 

Instead of executing the computed graph Wiener filter in the GFT domain, there exist fast methods based on Chebyshev polynomials \cite{hammond11} or Lanczos method \cite{susnjara15} so that processing can be carried out locally in the vertex domain.
Graph-signal filtering will be covered in more details in Section\;\ref{sec:filter}.

\subsubsection{Other Graph-based Image Denoising Approaches}

We overview a few other notable approaches in graph-based image denoising. 
\cite{tian14} performed image denoising by projecting an observed signal to a low-dimensional Krylov subspace of the graph Laplacian via a conjugate gradient method, resulting a fast image filtering operation that is competitive with Chebyshev polynomial approximation for the same order.
As an extension, \cite{knyazev15} performs edge sharpening using a graph with negative edges, implemented using the same projection method via conjugate gradient. 
\cite{gadde17} proposed a fast graph construction to mimic the performance of an edge-preserving bilateral filter (BF), where the computed sparse graph has eigenvectors in the graph spectral domain that are very close to the original BF. 
Edge-preserving smoothing is also considered in \cite{talebi16} via multiple Laplacians of affinity weights, each of which avoids computation-expensive normalization.

\subsection{Image Deblurring}

Image deblurring is more challenging than denoising, where the image model (\ref{eq:restoreModel}) has  a blurring operator $\mathbf{H}$, which may or may not be known. 
Among many proposals in the literature \cite{takeda08,xue13} is \cite{kheradmand14}, which elects a graph-based approach.
The unique aspect in \cite{kheradmand14} is that the similarity matrix $\mathbf{W}$ is first pre- and post-multiplied by a diagonal matrix $\mathbf{C}$, so that the resulting matrix $\mathbf{K}$ is both row- and column-stochastic:
\begin{equation}
\mathbf{K} = \mathbf{C}^{-1/2} \mathbf{W} \mathbf{C}^{-1/2}
\end{equation}
$\mathbf{C}$ is computed using a fast implementation \cite{knight12} of the Sinkhorn-Knop matrix scaling algorithm \cite{sinkhorn67}.
The resulting normalized Laplacian $\mathcal{L} = \mathbf{I} - \mathbf{K}$ is symmetric, positive semi-definite, and has the constant vector associated with eigenvalue $0$. 
This results in the following objective (\cite{kheradmand14}, eq.(16)):
\begin{align}
\min_{\mathbf{x}} &  \left( \mathbf{y} - \mathbf{H} \mathbf{x} \right)^T 
\left\{ \mathbf{I} + \beta (\mathbf{I} - \mathbf{K}) \right\} \left( \mathbf{y} - \mathbf{H} \mathbf{x} \right) \nonumber \\
& + \eta \mathbf{x}^{T} (\mathbf{I} - \mathbf{K}) \mathbf{x}
\label{eq:deblur}
\end{align}
where $\beta \geq -1$ and $\eta > 0$ are parameters.
Note that formulation (\ref{eq:deblur}) is useful for any linear inverse problems. 
The solution $\mathbf{x}^*$ of (\ref{eq:deblur}) can be obtained by solving a system of linear equations via conjugate gradient. 
Similarity matrix $\mathbf{W}$ is then updated using computed $\mathbf{x}^*$, and the process is repeated for several iteration to remove blur.

In another approach, \cite{yamamoto16} extends the SURE-LET image deblurring framework in \cite{xue13} for point cloud attributes (e.g., texture on 3D models). 
The key idea is to use graph to represent irregular 3D-point structures in a point cloud, so that subband decomposition and Wiener-like filtering via thresholding can be performed before reconstructing the signal. 
The blur kernel is replaced by Tikhonov regularized inverse for better condition number. 
See \cite{yamamoto16} for details.

\subsection{Soft Decoding of JPEG Encoded Images}
\label{subsec:softDecode}


The graph Laplacian regularizer, which promotes PWS behavior in the reconstructed signal when used iteratively \cite{hu13,pang17}, can be used in combination with other priors for image restoration; an earlier work \cite{zhai13,liu14tip} combined the graph Laplacian regularizer with a kernel method for image restoration. 
To illustrate how different priors can be combined, we discuss the problem of soft decoding of JPEG images \cite{liu17}.
JPEG remains the prevalent image compression format worldwide, and thus optimizing image reconstruction from the compressed format remains important. 
Recall that in JPEG, each $8 \times 8$ pixel block is transformed via DCT to coefficients $Y_i$, each of which is scalar quantized:
\begin{equation}
q_i = \textrm{round}\left( Y_i/Q_i \right)
\end{equation}
where $Q_i$ is the quantization parameter (QP) for coefficient $i$. 
The quantized coefficients of different blocks are subsequently entropy-coded into the JPEG compressed format.

At the decoder, one must decide which coefficient value $Y_i$ to reconstruct within the indexed quantization bin before inverse DCT to recover the pixel block:
\begin{equation}
q_i Q_i \leq Y_i < (q_i+1) Q_i
\label{eq:quanBin}
\end{equation}

To choose $Y_i$ within the bin constraint (\ref{eq:quanBin}), one must rely on signal priors. 
In \cite{liu17}, the authors used a combination of three priors that complement each other: Laplacian distribution for DCT coefficients, sparse representation given a compact pre-trained dictionary, and a new graph-signal smoothness prior. 
For initialization of the first solution, the first prior assumes that each DCT coefficient $i$ follows a Laplacian distribution with parameter $\mu_i$ \cite{lam00}.
The second prior assumes that a pixel patch can be approximated by a sparse linear combination $\boldsymbol{\alpha}$ of atoms from an over-complete dictionary $\boldsymbol{\Phi}$ \cite{elad06}.
\cite{liu17} shows that if $\boldsymbol{\Phi}$ is constrained in size due to computation cost, then the reconstructed patch would lack high DCT frequencies, resulting in blurs.

Finally, a new graph-signal smoothness prior using the \textit{Left Eigenvectors of Random walk Graph Laplcian} (LERaG) is proposed.
As previously discussed, iterative graph Laplacian regularizer promotes PWS behavior, thus recovering lost high DCT frequencies in a PWS pixel patch and complementing the restoration abilities of the aforementioned sparse coding using a small over-complete dictionary.

Further, for patch-based restoration, it is desirable in general to apply the \textit{same} filtering strength when processing different patches in the image. 
Using previously described regularizer $\mathbf{x}^T \mathbf{L} \mathbf{x}$ where $\mathbf{L}$ is unnormalized, however, would mean that the strength of the resulting filtering depends on the total degree of the constructed graph. 
One alternative is to use the symmetric normalized graph Laplacian $\mathcal{L}$ to define smoothness prior $\mathbf{x}^T \mathcal{L} \mathbf{x}$.
However, because the constant signal is not an eigenvector of $\mathcal{L}$, prior $\mathbf{1}^T \mathcal{L} \mathbf{1} > 0$, and the prior does not preserve constant signals that are common in natural images.

\begin{figure}[htb]
\begin{minipage}[b]{.32\columnwidth}
  \centering
  \centerline{\includegraphics[width=3.0cm]{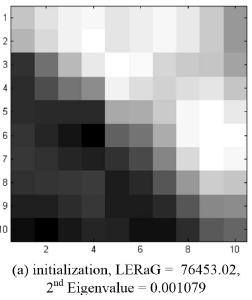}}
\end{minipage}
\hfill
\begin{minipage}[b]{.32\columnwidth}
  \centering
  \centerline{\includegraphics[width=3.0cm]{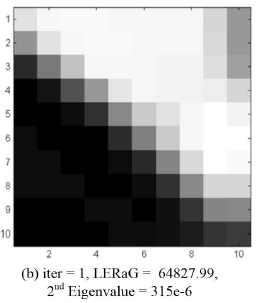}}
\end{minipage}
\hfill
\begin{minipage}[b]{.32\columnwidth}
  \centering
  \centerline{\includegraphics[width=3.0cm]{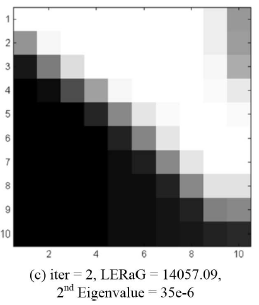}}
\end{minipage}
\vspace{-0.1in}
\caption{2nd eigenvalue of normalized graph Laplacian is monotonically decreasing with iteration numbers.}
\label{fig:3-lerag}
\end{figure}

Instead, \cite{liu17} proposed to use $\mathbf{x}^T \mathcal{L}_r^T \mathcal{L}_r \mathbf{x}$, computed efficiently as $\mathbf{x}^T (d^{-1}_{\min}) \mathbf{L} \mathbf{D}^{-1} \mathbf{L} \mathbf{x}$, as the graph-signal smoothness prior, where $\mathcal{L}_r = \mathbf{D}^{-1} \mathbf{L}$ is the random walk graph Laplacian matrix. 
Like $\mathcal{L}$, $\mathcal{L}_r$ is normalized so the filter strength of the derived processing is the same for different patches. 
Yet unlike $\mathcal{L}$, $\mathbf{1}^T \mathcal{L}_r^T \mathcal{L}_r \mathbf{1} = 0$, and hence the prior can well preserve constant signals in natural images.
Compared to the new normalized graph Laplacian matrix computed from a doubly stochastic similarity matrix as discussed earlier \cite{kheradmand14}, \cite{liu17} showed that LERaG outperforms this approach with a lower computation cost (see \cite{liu17} for detailed comparisons).

Combining the sparsity prior and LERaG, we arrive at the following optimization for soft decoding of a pixel patch $\mathbf{x}$:
\begin{equation}
\label{eq:obj_priors}
\begin{array}{l}
\mathop{\arg \min} \limits_{\{\mathbf{x}, \boldsymbol{\alpha}\}}  
\left\| 
{\mathbf{x} - \boldsymbol{\Phi} \boldsymbol{\alpha} } \right\|_2^2 + {\lambda_1}{\left\| \boldsymbol{\alpha}  \right\|_0} + \lambda_2\x^T (d_{\min}^{-1})\mathbf{L} \mathbf{D}^{-1} \mathbf{L}\mathbf{x}, \\
\mbox{s.t.}~ \mathbf{q}\mathbf{Q} \preceq \mathbf{T}\mathbf{M}\mathbf{x} \prec (\mathbf{q}+1)\mathbf{Q}
\end{array}
\end{equation}
where $\lambda_1$ and $\lambda_2$ are weight parameters, $\mathbf{q}$ and $\mathbf{Q}$ are the quantization bin indices and QP's, and both signal $\mathbf{x}$ and its sparse code $\boldsymbol{\alpha}$ are unknown.
$\mathbf{x}$ and $\boldsymbol{\alpha}$ are solved alternately while holding the other variable fixed.

\begin{figure}[ht!]
\centering
\includegraphics[width = 0.49\textwidth]{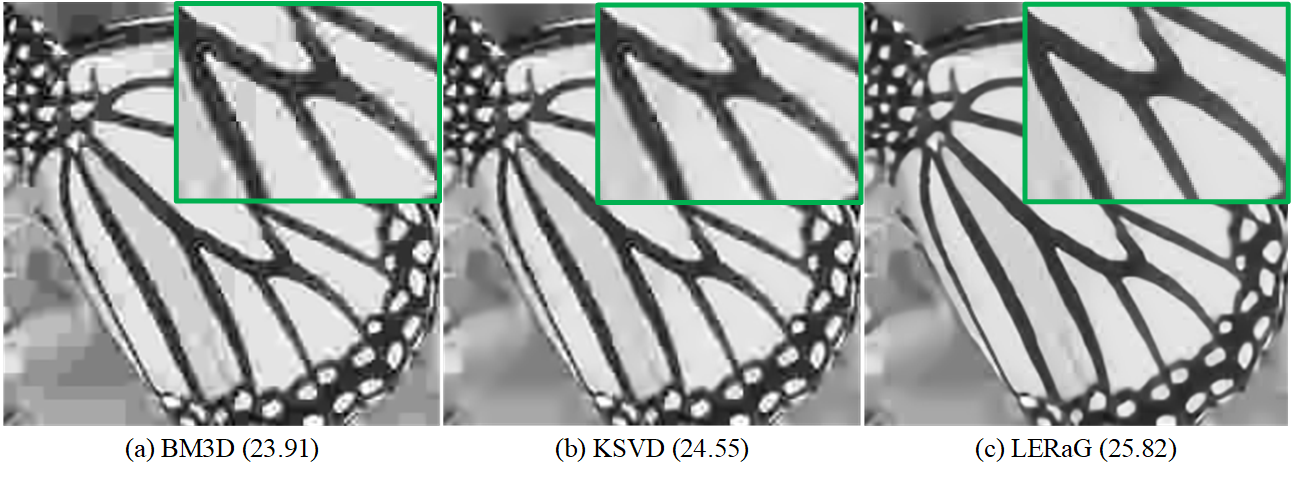}
\caption{Comparison of tested methods in visual quality on \emph{Butterfly} at QF = 5. The corresponding PSNR values are also given as references.}
\label{fig:3-butterfly}
\vspace{-0.1cm}
\end{figure}

As suggested in \cite{milanfar13}, to improve filtering performance, (\ref{eq:obj_priors}) is computed iteratively, each time the edge weights in the graph are updated from the last computed solution $\mathbf{x}$. 
Due to the diffusion taking place as discussed previously, the filtered patch will increasingly become more PWS, as shown in Fig.\;\ref{fig:3-lerag}.
Note also that the second eigenvalue of the normalized graph Laplacian becomes increasingly smaller, resulting in a smaller prior cost. 
As shown in Fig.\;\ref{fig:3-butterfly}, the soft-decoded JPEG image \texttt{Butterfly} has higher quality than competing schemes. 

\subsection{Other Graph-based Image Restorations}
\label{subsec:otherRestore}

To show the breadth of applications using graph-based image restoration techniques, we briefly overview a few notable works. 
\cite{hu12icip2} first interpolates a full color image from Bayer-patterned samples, then based on the interpolated values computes edge weights for graph Laplacian regularization towards image demosaicking.
For a stereo image pair with heterogeneous qualities, the higher-quality view image and corresponding disparity map are used to construct appropriate graph for bilateral filtering of the lower-quality view image in order to suppress noise \cite{tian14icip}.
Similar in concept, leveraging on the information provided by the high-resolution color image, resolution of the low-resolution depth image is enhanced via joint bilateral upsampling \cite{wang14}.
An image bit-depth enhancement scheme using a graph-signal smoothness assumption for the AC component in an image patch was proposed in \cite{wan16}.

\section{Graph-based Image Filtering}
\label{sec:filter}
As in image denoising and other inverse imaging, extracting smooth components of the image, i.e., low-graph-frequency components, is a critical issue since many image filtering applications utilize edge-preserving image smoothing as a key ingredient. This section introduces various image filtering methods using graph spectral analysis and shows relationships among them.


\subsection{Smoothing and Diffusion in Graph Spectral Domain}

One of the seminal works on smoothing using graph spectral analysis is 3-D mesh processing from computer graphics community \cite{taubin95, taubin96}\footnote{The term ``graph signal" was first introduced in \cite{taubin96}, to the best of our knowledge.}. It determines edge weights of the graph as Euclidean distance between vertices (of the 3-D mesh) and smooths the 3-D mesh shape using a graph low-pass filter with a binary response. That is, the spectral response of the filter is
\begin{equation}
\label{eqn:taubinfilter}
\widetilde{h}(\lambda_k) = \begin{cases}
1      & \text{if } k \le T_k, \\
0      & \text{otherwise},
\end{cases}
\end{equation}
where $T_k$ is the user-defined bandwidth, i.e., how many eigenvalues are passed. Clearly, we can define an arbitrary response according to the purpose. This kind of naive approaches have been used in several computer graphics/vision tasks \cite{zhang10, vallet08, desbrun99, fleishman03, kim05}. The filter in \eqref{eqn:taubinfilter} actually smoothes out high-graph-frequency components, however, as the number of vertices grows, it is difficult to compute graph Fourier basis via eigendecomposition.

Heat kernel in the spectral domain has also been proposed in \cite{zhang08}. In this work, the weight of the edges of the graph is computed according to photometric distance, i.e., large weights are assigned to the edges whose both ends have similar pixel values and vice versa. Additionally, its graph spectral filter is defined as a solution of the heat equation on the graph as follows:
\begin{equation}
\label{eqn:heatkernel}
\widetilde{h}(\lambda) = e^{-t\lambda},
\end{equation}
where $t > 0$ is an arbitrary parameter to control the spreading speed due to diffusion. By implementing it with the naive approach, it still needs a large computation cost due to eigendecomposition of graph Laplacian. However, \eqref{eqn:heatkernel} can also be represented by using Taylor series around the origin as
\begin{equation}
\label{eqn:heat_taylor}
e^{-t\lambda} = \sum_{k=0}^{\infty} \frac{t^k}{k!} (-\lambda)^k.
\end{equation}
By truncating the above equation with an arbitrary order, we can approximate it as a finite-order polynomial \cite{hammond11, shuman13}. In \cite{zhang08}, the Krylov subspace method is used along with \eqref{eqn:heat_taylor} to approximate the graph filter.

\begin{figure}[t]
\centering
\includegraphics[width=.8\linewidth]{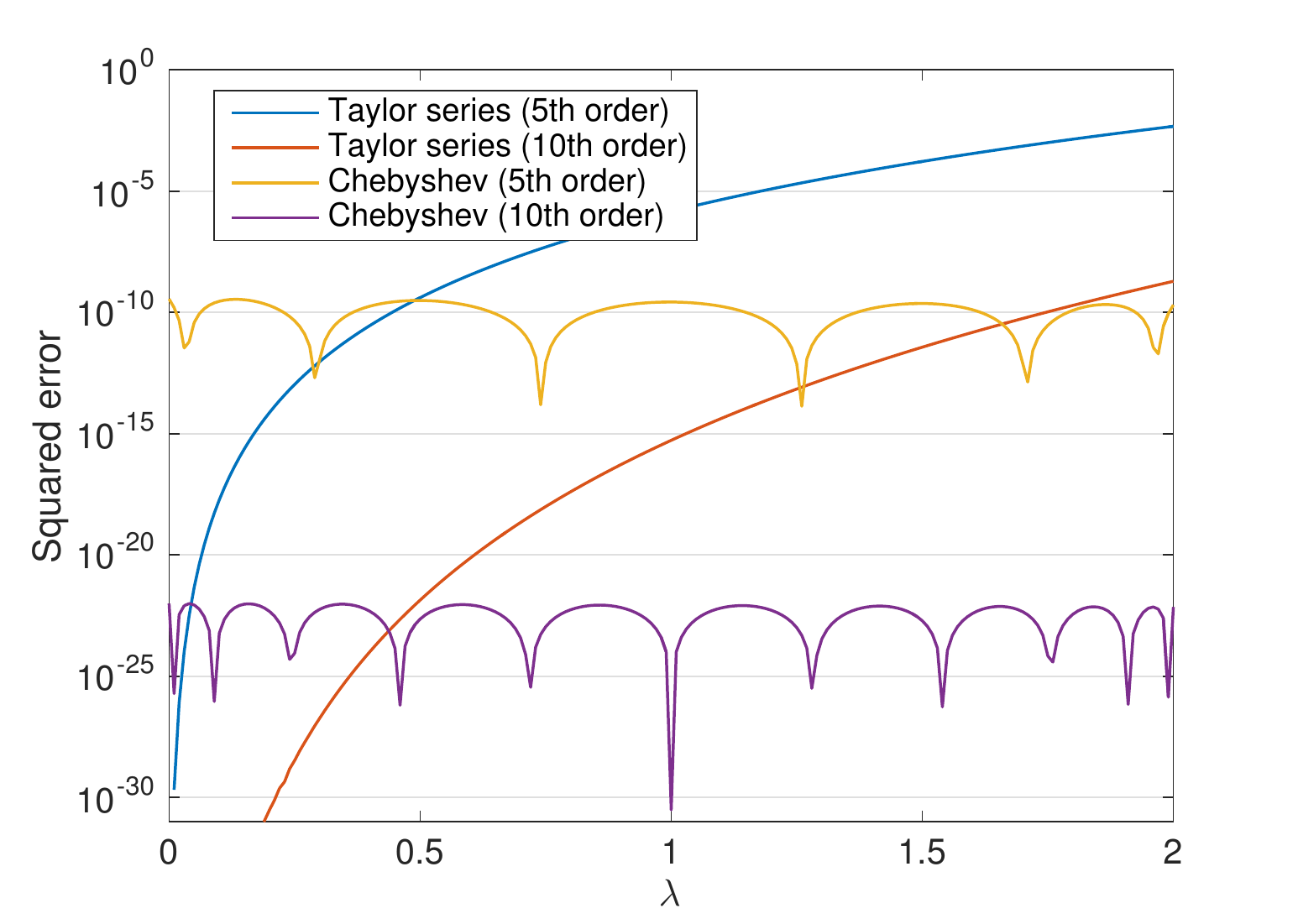}
\caption{Approximation error comparison of $\widetilde{h}(\lambda) = e^{-\lambda}$. The error is calculated as $E(\lambda) = (\widetilde{h}(\lambda)-\widetilde{h}_K^{\text{approx}}(\lambda))^2$, where $\widetilde{h}_K^{\text{approx}}(\lambda)$ is the $K$th-order approximated response.}
\label{fig:approxerror}
\end{figure}

However, as shown in Fig. \ref{fig:approxerror}, its approximation accuracy significantly gets worse for large $\lambda$. Since the maximum eigenvalue $\lambda_{\max}$ highly depends on the graph used, it is better to use different approximation methods (introduced in Section \ref{subsec:fastcomp}).

\subsection{Edge-Preserving Smoothing}
As previously mentioned, edge-preserving smoothing is widely used for various image filtering tasks as well as image restoration \cite{nagao79, pomalaza-raez84, weickert98, tomasi98, barash02, durand02, farbman08, xu11, he13}. Image restoration aims to get the ground-truth image (approximately) from its degraded version, whereas edge-preserving smoothing is used to yield a user-desired image from the original one; It is either noisy or noise-free.

In the graph setting, we often need to define pixel-wise or patch-wise relationships as a distance between pixels or patches, and it is used to construct a graph. Three distances are considered in general \cite{milanfar13}: 1) geometric distance, 2) photometric distance, and 3) these combination.
Furthermore, especially for image filtering other than restoration, we often employ 4) \textit{saliency} of the image/region/pixel, which simulates perceptual behavior \cite{itti98, harel06}.

The graph spectral representation of bilateral filter \cite{gadde13} introduces that the bilateral filter can be regarded as a combination of  graph Fourier basis and a graph low-pass filter. The filter coefficients of the bilateral filter is represented in \eqref{eq:edgeWeight}.
Since its weights clearly depend on the geometric and photometric distances, it is a pixel-dependent filter. 

In a classical sense, the frequency domain representation of the bilateral filter cannot be calculated straightforwardly. In contrast, the bilateral filter can be considered as a graph filter by considering a weight matrix $\mathbf{W}$ where $[\mathbf{W}]_{ij} = w_{i,j}$ as an adjacency matrix of the graph. \eqref{eq:edgeWeight} is rewritten as
\begin{equation}
\label{ }
\widehat{\mathbf{x}} = \mathbf{D}^{-1}\mathbf{W} \mathbf{x}
\end{equation}
where $\mathbf{D} = \text{diag}(d_0, d_1, \ldots, d_{N-1})$ in which $d_i = \sum_j w_{i,j}$. It is further rewritten as a graph spectral filter by
\begin{equation}
\label{ }
\widehat{\mathbf{x}} = \mathbf{D}^{-1/2}\mathbf{U}_n(\mathbf{I} -\mathbf{\Lambda}_n)\mathbf{U}^T_n\mathbf{D}^{1/2}\mathbf{x}
\end{equation}
%
where we utilize the fact $\mathbf{W} = \mathbf{D} -\mathbf{L}$ and $\mathbf{L}_n = \mathbf{D}^{-1/2}\mathbf{L}\mathbf{D}^{-1/2}$. When we define a degree-normalized signal as $\underline{\mathbf{x}} = \mathbf{D}^{-1/2}\mathbf{x}$, the above equation is represented as
\begin{equation}
\label{eqn:graphBF}
\underline{\widehat{\mathbf{x}}} = \mathbf{U}_n\widetilde{h}_\text{BF}(\mathbf{\Lambda}_n)\mathbf{U}^T_n\underline{\mathbf{x}},
\end{equation}
where $\widetilde{h}_\text{BF}(\lambda_n) = 1 - \lambda_n$. Since $\lambda_n \in [0, 2]$, it acts as a graph low-pass filter.

The above-mentioned representation of the bilateral filter suggests that the original bilateral filter implicitly designs the graph Fourier basis and the graph spectral filter simultaneously. For example, consider the following spectral response:
\begin{equation}
\label{eqn:graphfilter_denoising}
\widetilde{h}(\lambda) = \frac{1}{1 + \rho \widetilde{h}_r(\lambda)},
\end{equation}
where $\widetilde{h}_r(\lambda)$ is a graph high-pass filter and $\rho > 0$ is a parameter. Clearly $\widetilde{h}(\lambda)$ works as a graph low-pass filter. It is the optimal solution of the following denoising problem \cite{gadde13}:
\begin{equation}
\label{ }
\argmin_{\underline{\mathbf{x}}} ||\underline{\mathbf{y}} - \underline{\mathbf{x}}||_2^2 + \rho||\mathbf{H}_r  \underline{\mathbf{x}}||_2^2,
\end{equation}
where $\mathbf{y} = \mathbf{x} + \mathbf{e}$ in which $\mathbf{e}$ is a zero-mean i.i.d. Gaussian noise and $\mathbf{H}_r = \mathbf{U} \widetilde{h}_r(\mathbf{\Lambda})\mathbf{U}^T$.

It is known that the bilateral filter sometimes needs many iterations to smooth out details for textured and/or noisy images. To boost up the smoothing effect, the trilateral filter method \cite{choudhury03} first smoothes gradients of the image, and then the smoothed gradient is utilized to smooth intensities. Its counterpart in the graph spectral domain has also proposed in \cite{onuki16} with the parameter optimization method for $\rho$ in \eqref{eqn:graphfilter_denoising} which minimizes MSE after denoising.

Other than the bilateral filter, non-local filters can also be interpreted as graph spectral filters like \eqref{eqn:graphBF} while variational operators are not restricted to the symmetric normalized graph Laplacian and sometimes they are permitted to have negative-weighted edges. For example, \cite{talebi16, talebi14} introduce graph spectral filters based on non-local means \cite{buades05a} with random-walk graph Laplacian.

The power of graph spectral analysis for image filtering is that it is able to consider the prior information of the image, e.g., edges, textures, and saliencies, as a graph, separately from the user-desired information as a graph spectral response. That is, we can (and need to) design good graphs as well as graph filters for the desired image filtering effects. The design methods of graph spectral filters for various image processing tasks have been discussed in \cite{talebi14, talebi16, talebi14a, rong08, zhang10, yagyu16} and references therein.

The above approach is generally represented as
\begin{equation}
\label{eqn:psdfilter}
\widehat{\mathbf{x}} = \mathbf{W}\widetilde{h}(\mathbf{\Lambda})\mathbf{W}^{-1}\mathbf{x},
\end{equation}
where $\mathbf{W} \in \mathbb{R}^{N \times N}$ is an arbitrary dictionary which sparsely represents the image $\mathbf{x}$ \footnote{Generally the number of atoms in the dictionary could be overcomplete, but we focus on the square and invertible $\mathbf{W}$ for the sake of simplicity.} and $\widetilde{h}(\mathbf{\Lambda}) = \text{diag}(\widetilde{h}(\lambda_0), \widetilde{h}(\lambda_1), \ldots)$ is a filter in the spectral domain (not restricted to the spectrum of the graph). This general form is considered in a modern image processing tasks \cite{milanfar13} where $\mathbf{W}$ and $\mathbf{\Lambda}$ are obtained from a (symmetrized) matrix whose elements are come from arbitrary image processing. However, in this paper, we focus on a specific form of \eqref{eqn:psdfilter} in graph setting, where the dictionary is so-called graph Fourier basis and the spectral response is designed for the eigenvalues of the graph.

\subsection{Relationship between Edge-Preserving Smoothing and Retargeting}
For various image processing tasks including the topics described in this paper, filtering methods combining geometric and photometric distances and saliency have been proposed (see \cite{tomasi98, buades05, buades05a, takeda07, bhat10, paris11, milanfar13, he13, aubry14} and references therein). Among the works, \textit{domain transform} \cite{gastal11, gastal15} has a unique approach: It transforms the photometric distance  into the geometric distance, then the nonuniformly distributed discrete signal is low-pass filtered. Finally, the nonuniformly distributed signal is warped to its original pixel position to obtain the resulting smoothed image.

Formally, the domain transform for a 1-D signal is performed in the following steps.
\begin{enumerate}
  \item Compute a warped pixel position according to the geometric and photometric distances.
\begin{equation}
\label{ }
t_i  = t_{i-1} + \alpha_g + \alpha_p \sum_{k=0}^{N_c - 1} |x^{(k)}_i - x^{(k)}_{i-1}|,
\end{equation}
where $t_i$ is the $i$th pixel position ($t_0 = 0$), $\alpha_g$ and $\alpha_p$ are the weights for the geometric and photometric distances (usually $\alpha_g = 1$), respectively, $x^{(k)}_i$ is the $i$th pixel value of the $k$th color component, and $N_c$ is the number of color channels, e.g., $N_c = 3$ for RGB color images.
  \item Place $x_i$ onto $t_i$ as $f(t_i) := x_i$. At this time, $f(t_i)$ can be regarded as a nonuniformly sampled continuous signal.
  \item Perform a low-pass filter $h(t)$ to $f(t_i)$ to obtain $\widehat{f}(t) = h(t) \ast f(t)$ defined in the continuous domain $t \in [0, t_{N-1}]$.
  \item Replace the filtered signal $\widehat{f}(t_i)$ back to its original coordinates, i.e., $\widehat{x}_i = \widehat{f}(t_i)$.  
\end{enumerate}
The motivation of the domain transform is clearly shared with that of the graph-based image processing; The relationship between signal values is determined first, then the low-pass filter is performed to obtain user-desired effects.

The deformed pixel position $t_i$ can also be regarded as a solution of the following linear problem \cite{yagyu16}.
\begin{equation}
\label{eqn:domaintransmatrix}
\mathbf{\Psi} \mathbf{t} = \bm{\tau},
\end{equation}
where
\begin{equation}
\label{}
[\mathbf{\Psi}]_{ij} = \begin{cases}
1      & i = j, \\
-1 &  i = j-1,\\
0     & \text{otherwise}
\end{cases}
\end{equation}
and $\tau_i = \alpha_g + \alpha_p \sum_{k=0}^{N_c - 1} |x^{(k)}_{i} - x^{(k)}_{i-1}|$. \eqref{eqn:domaintransmatrix} can be solved by simply taking the inverse of $\mathbf{\Psi}$ and it is represented as
\begin{equation}
\label{ }
\mathbf{\Psi}^{-1} = \begin{bmatrix}
1 & 0 & \cdots & \\
1 & 1 & \ddots & \\
\vdots & \ddots & \ddots \\
\end{bmatrix}.
\end{equation}

Here, let us consider the following optimization problem:
\begin{equation}
\label{}
\argmin_{\mathbf{t}}||\bm{\tau} - \mathbf{\Psi}\mathbf{t}||^2_2.
\end{equation}
Its solution is $\mathbf{\Psi}^T\mathbf{\Psi} \mathbf{t} = \mathbf{\Psi}^T\bm{\tau}$ and \eqref{eqn:domaintransmatrix} is obtained when we multiply $\mathbf{\Psi}^{-T}$ for both sides. Interestingly, the above linear problem can also be represented as
\begin{equation}
\label{eqn:Lpathx}
\mathbf{L}_{\text{path}} \mathbf{t} - \bm{\tau}' = \mathbf{0},
\end{equation}
where $\mathbf{L}_{\text{path}} = \mathbf{\Psi}^T\mathbf{\Psi}$ is the graph Laplacian of the path graph and $\tau'_i = \alpha_p \sum_{k=0}^{N_c - 1} |x^{(k)}_{i+1} - x^{(k)}_{i}| - |x^{(k)}_{i} - x^{(k)}_{i-1}|$. It means we can define a generalized version of the domain transform and the general form is very closely related to various mesh deformation methods \cite{gu02, desbrun02, zhang10, rong08, wang08}.

Mesh deformation is widely used in computer graphics and vision as well as image processing. The simplest form of the optimization problem can be formulated as follows \cite{sorkine04, zhou05, wang08}:
\begin{equation}
\label{eqn:meshdeform}
\mathbf{L}_1 \mathbf{p}' - \mathbf{L}_0 \mathbf{p} = \mathbf{0}
\end{equation}
where $\mathbf{L}_0$ and $\mathbf{L}_1$ are, respectively, the graph Laplacians for the original and deformed vertices of the mesh and $\mathbf{p}$ and $\mathbf{p}'$ are the original and deformed vertex coordinates, respectively. Since a pure Laplacian is a singular matrix, \eqref{eqn:meshdeform} needs constraints to obtain robust solutions. One of the widely-used constraints is the boundary condition which keeps the deformed vertex positions on the boundary unchanged.

\eqref{eqn:Lpathx} is a special version of \eqref{eqn:meshdeform} since $\bm{\tau}'$ in \eqref{eqn:Lpathx} represents the second-order differentiation of the deformed pixel position. Conversely, if we can define a ``good" distance (depending on applications) between a pair of pixels, its deformed pixel position would be determined by solving a linear equation having the form like \eqref{eqn:meshdeform}, as long as the overall cost function is quadratic. The desired pixel values could be obtained by a graph-based filtering.

\begin{figure*}[t]
\centering
\includegraphics[width=.9\linewidth]{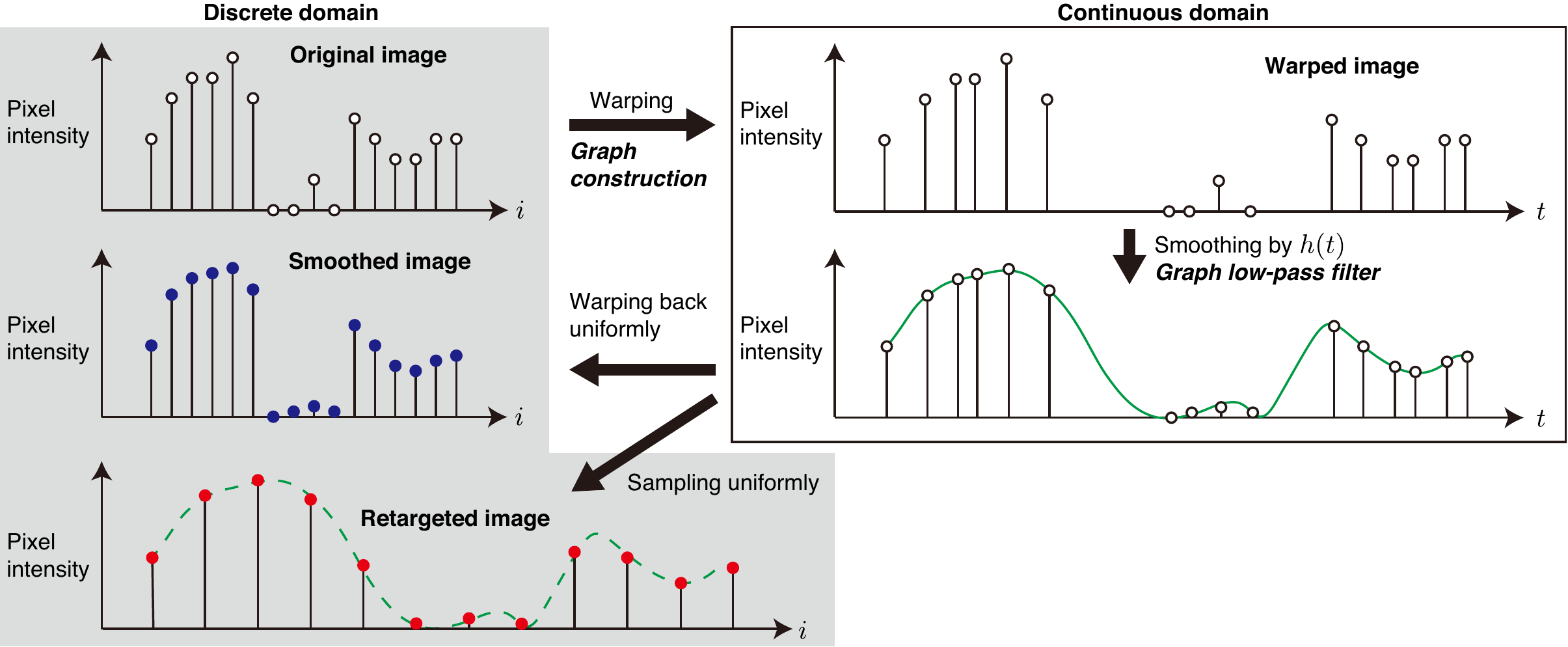}
\caption{Relationship between edge-preserving smoothing and image retargeting as signal processing on nonuniform grid. Graph signal processing corresponds to the construction of the appropriate graph and filtering of the nonuniformly sampled signal.}
\label{fig:smoothing_retargeting}
\end{figure*}

If we perform a low-pass filter to the nonuniformly distributed pixel $f(t'_i)$ then move it back to its original uniform-interval position, the graph-based filtering is an edge-preserving smoothing. Instead of that, if we interpolate the uniform-interval pixels $\widehat{f}(t_i)$ from $f(t'_i)$, it is so-called content-aware image resizing, also known as \textit{image retargeting} \cite{wolf07, wang08, zhang09shape}. Their relationship is illustrated in Fig. \ref{fig:smoothing_retargeting}.

Note that the conventional approaches need to consider signal processing in a continuous domain as a counterpart of the discrete domain where image signals exist. It leads to that we have to estimate the appropriate continuous domain from the input signal or any other prior information. Generally it is a difficult task and requires a large computation cost due to signal processing in the continuous domain. In contrast, graph-based methods are fully discretized; memory and computation costs are usually kept low. Additionally, the prior information is appropriately utilized to construct/learn a graph for purposes.

\subsection{Non-Photorealistic Rendering of Images}
\begin{figure}[t]
\centering
\subfigure{\includegraphics[width=.53\linewidth]{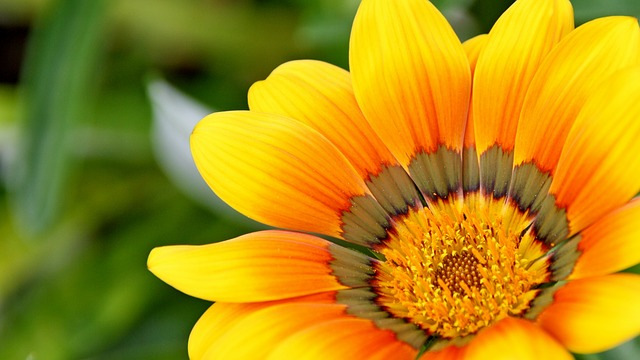}}\ \subfigure{\includegraphics[width=.45\linewidth]{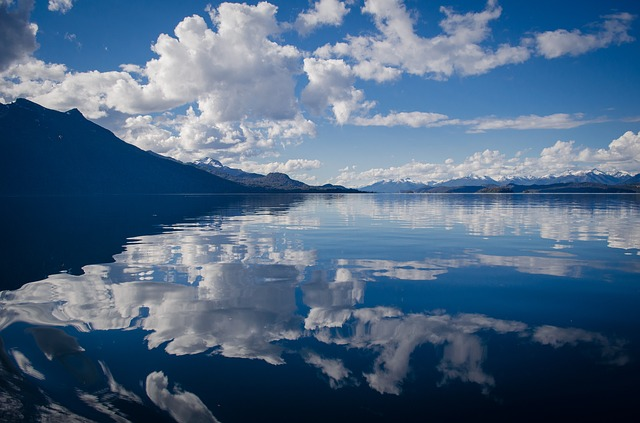}}
\subfigure{\includegraphics[width=.53\linewidth]{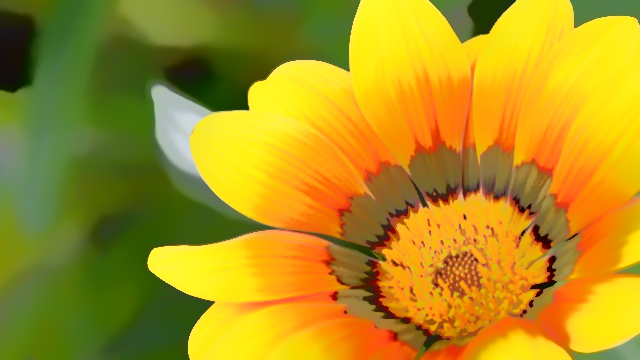}}\ \subfigure{\includegraphics[width=.45\linewidth]{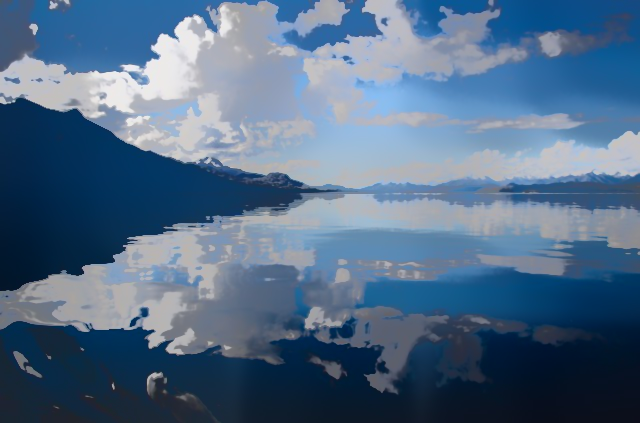}} 
\subfigure{\includegraphics[width=.53\linewidth]{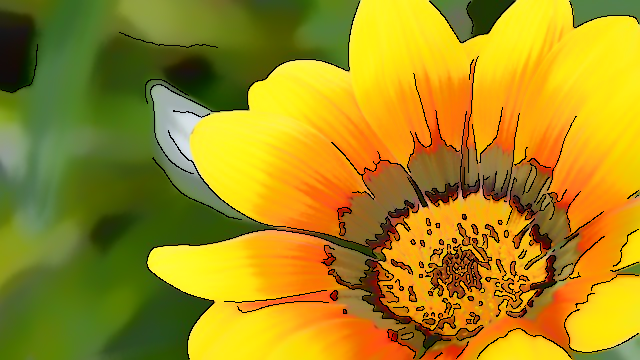}}\ \subfigure{\includegraphics[width=.45\linewidth]{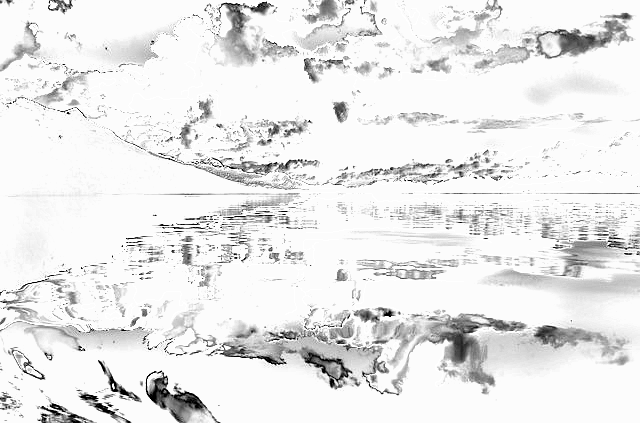}}
\caption{Examples of non-photorealistic rendering. Image stylization is shown in the left column and pencil drawing is shown in the right column. From top to bottom: Original, edge-preserving smoothing, and NPR results, respectively. The method in \cite{talebi14a} is used for edge-preserving smoothing. For stylization, the edge image is combined with the smoothed image. For pencil drawing, edge detection is performed to the smoothed image and the edge image is combined with the high-frequency information in the image. Both test images are obtained from \protect\url{https://pixabay.com/}.}
\label{fig:npr_example}
\end{figure}

Edge-preserving smoothing is also widely used in non-photorealistic rendering (NPR), which is one of key tasks in computer graphics. With a combination of image processing techniques like thresholding (both in spatial and frequency domains) and segmentation, NPR accomplishes various artificial effects such as stylization, pencil drawing, and abstraction \cite{winkenbach94, strothotte02, rusinkiewicz08, gastal11, yagyu16, yagyu16a, sadreazami17}. Examples of NPR are shown in Fig. \ref{fig:npr_example}.

Sometimes one needs NPR images with different degrees of artificiality. We can accomplish them by defining different graphs and filters for different artificialities, however, it is generally a cumbersome process. Instead, multiscale decomposition of images would be an alternative way. Traditionally, each scale represents an image component which has a specific frequency range. In contrast, graph-based multiresolution has more flexibility on the preserved component in each scale; It can also reflect the structure of pixels in each scale. For example, when we apply graph Laplacian pyramid \cite{shuman16} to an image with an appropriate multiscale graph and graph filters, high-graph-frequency component could represent pixel-level fine details, while the upper (coarser) level component would have region-level salient features. By changing functions to strengthen and weaken the transformed coefficients in each scale, we can obtain different NPR results from one multiscale representation of images. The multiscale representation has been proposed in the literature \cite{rong08, zhang10, yagyu16}.

\subsection{Fast Computation}
\label{subsec:fastcomp}

Fast computation of graph spectral filtering is a key for its practical applications since the modern image filtering, including graph spectral filters, treats image signals as one long-vector $\mathbf{x} \in \mathbb{R}^{WH}$ where $W$ and $H$ are the width and height of the original image, respectively. Image resolution of digital broadcasting is becoming larger and larger; For example, 4K ultra-high-definition corresponds to $W = 3840$ and $H = 2160$ pixels, which leads to $WH > 8 \times 10^{6}$. As the naive approach, we have to construct a graph Laplacian of the size $WH \times WH$, then perform its eigendecomposition. This approach needs huge computational burden even in recent high-spec computers, and therefore, a workaround should basically be considered.

Although there are various methods to realize approximate spectral graph filtering, they can be divided into two approaches. One uses approximated eigenvectors (and eigenvalues) and the exact spectral filter response. The other uses an approximated spectral filter response and exact eigenvectors.

The first approach computes eigenvectors (or singular vectors for rectangular matrices) partially and/or approximately. Remaining eigenvectors are often approximated from the calculated eigenvectors. This approach can further be classified into two categories: Computing approximate eigenvectors from 1) graph Laplacian or other variation operators \cite{fowlkes04, drineas05, oh15}, and 2) pre-filtered images \cite{talebi14, talebi14a}. Both can be applicable to any real symmetric matrices and the Nystr{\"o}m approximation method \cite{william01} plays a central role. They can drastically reduce the computation cost whereas it is required to decide how many eigenvectors are calculated prior to the decomposition.


The second approach uses the spectral response represented as a polynomial \cite{saad84, kokiopoulou04, zhou07, cai13, talebi16, onuki17}. Generally, if the filter response in the graph spectral domain is a $K$th order polynomial function $\widetilde{h}(\lambda) = \sum_{k=0}^{K-1} a_k \lambda^k$, it can be represented as $K$ matrix-vector multiplications \cite{shuman13} since
\begin{equation}
\label{ }
\mathbf{U}\widetilde{h}(\mathbf{\Lambda})\mathbf{U}^T = \mathbf{U}\left(\sum_{k=0}^{K-1} a_k \mathbf{\Lambda}^k \right)\mathbf{U}^T = \sum_{k=0}^{K-1} a_k \mathbf{L}^k,
\end{equation}
where we utilize the fact $\mathbf{L}^k = \mathbf{U}\mathbf{\Lambda}^k\mathbf{U}^T$. This means we can use the exact full eigenvectors for filtering while the spectral response is approximated. It leads to that we have to choose good $\{a_k\}$. They can be determined empirically according to desired filtering effect \cite{talebi14a, talebi16} or from polynomial approximation of the desired spectral response \cite{saad84, kokiopoulou04, zhou07, cai13, onuki17}.

The $K$th order polynomial function on the graph can also be represented as the $K$-hop neighborhood transform in the vertex domain \cite{shuman13}. It leads to the polynomial filters are localized in the vertex domain. Additionally, the output signal can be obtained from a distributed calculation.

Among the polynomial approximation methods, Chebyshev polynomial approximation \cite{mason02, phillips03, hammond11, shuman11, onuki17} is widely used for graph signal processing from some reasons. First, it can calculate with a recurrence relation; Memory requirement is small. Second, it produces low errors for the passband region. Third, it is very close to minimax polynomial and error bound can be calculated. Its approximation error for $\widetilde{h}(\lambda) = e^{-\lambda}$ is shown in Fig. \ref{fig:approxerror}. As compared with the Taylor series, the error by the Chebyshev approximation is bounded for all range of $\lambda$.

\section{Graph-based Image Segmentation}
\label{sec:segment}
Image segmentation is an important and fundamental step in
computer vision, image analysis and recognition \cite{gonzalez18}. It refers to
partitioning an image into different regions where each region has
its own meaning or characteristic in the image (e.g., the same color, intensity
or texture). In the literature, there are a large number of image segmentation methods including
threshold-based, edge-based, region-based and energy-based approaches; see the references in \cite{peng13}.
They have been applied to many image processing applications successfully, for example, in medical imaging,
tracking and recognition.

For image segmentation methods,
the energy-based approach is to develop and study
an energy function which gives an optimum when the image is segmented
into several regions according to the objective function criteria. This approach includes
several techniques such as active contour (e.g., \cite{kass88}) and graph cut (e.g., \cite{shi00,boykov01}).
The main advantage of using graph cut is that the associated energy function can be globally optimized whereas the
other segmentation methods may not be guaranteed.
In the graph cut segmentation, the energy function is constructed based on graphs where image pixels are
mapped to graph vertices, and it can be optimized via graph-based algorithms and spectral graph theory results.
By using the representation of graphs, morphological processing techniques can be applied to obtain many interesting image
segmentation results, see for instance \cite{najman14}.
In this paper, we focus on the concept of graph cut segmentation and discuss its application to
Mumford-Shah segmentation model.

\subsection{Graph Cut}

\begin{figure} \label{gc}

\vspace{-0.4cm}
\centerline{\includegraphics[width =3.8in, height=2.2in]{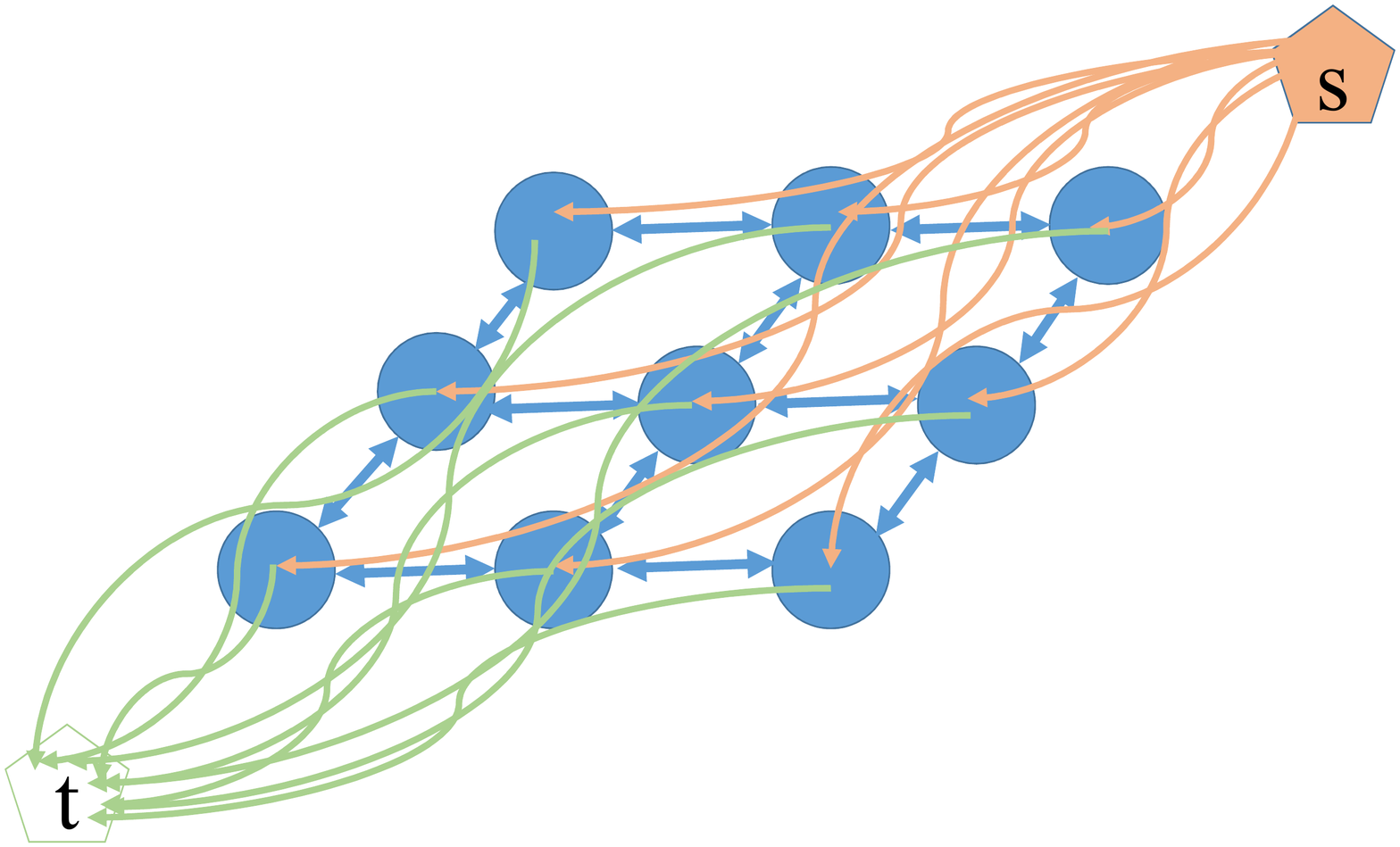}}
\vspace{-0.7cm}
\caption{An example of 3-by-3 image grid (blue circle: image pixel; blue arrowed line: pixel edge; brown arrowed line:
an edge from the source vertex to pixel vertex;
green arrowed line: an edge from pixel vertex to the sink vertex.}
\end{figure}

Given a graph ${\cal G} = ( {\cal V}, {\cal E} )$ composed of the vertex set ${\cal V}$ and
the edge set ${\cal E} \subset {\cal V} \times {\cal V}$. The vertex set ${\cal V}$ contains
the nodes of a two-dimensional or three-dimensional image pixels together with two terminal vertices:
the source vertex $s$ and the sink vertex $t$. The edge set ${\cal E}$ contains two kinds of edges: (i)
the edges $e = (i,j)$ where $i$ and $j$ are the image pixels except the source and the sink vertices;
(ii) the terminal edges $e_s = (s,i)$ and $e_t = (i,t)$ where $i$ is the image pixel except the source and the sink vertices.
In two-dimensional or three-dimensional images, we usually assign an edge between two neighborhood pixels. We refer
to Figure \ref{gc} for a 3-by-3 for illustration.
Moreover, the nonnegative cost $w_{i,j}$ is assigned to each edge $(i,j) \in {\cal E}$.

A cut on a graph is a partitioning of the vertices ${\cal V}$ into two disjoint and connected (through edges)
sets $({\cal V}_s,{\cal V}_t)$ such that $s \in {\cal V}_s$ and $t \in {\cal V}_t$.
For each cut, the set of served edges ${\cal C}$ is defined as follows:
$$
{\cal C}({\cal V}_s,{\cal V}_t) = \{ (i,j) \ | \ i \in {\cal V}_s, j \in {\cal V}_t \ {\rm and} \ (i,j) \in {\cal E} \}
$$
We say that the graph cut uses the served edge $(i,j)$ if $(i,j)$ is contained in ${\cal C}$.
Correspondingly, the cost of the cut is defined as follows:
$$
{\rm cost}({\cal C}({\cal V}_s,{\cal V}_t)) = \sum_{(i,j) \in {\cal C}({\cal V}_s,{\cal V}_t)} w_{i,j}
$$
In image segmentation, a cost function usually consists of the two terms: the region term and the boundary term \cite{boykov01}.
The region term is used to give a cost function for a pixel assigned to a specific region. For example,
the penalty can be referred to the difference between the intensity value of a pixel and the intensity model
of the region.
This term is usually used for the cost of edges between the source/sink vertex and pixel vertices.
The boundary term is used to give a cost function when two neighborhood pixels are assigned to
two different regions \cite{shi00}.
This term is usually used for the cost of edges between neighborhood pixels.

Basically, regional and edge information are used in graph cut.
By incorporating shape information of the object into graph cut, image segmentation results can be improved.
The main idea is to revise the region term and the boundary term in cost function such that specific image segmentation results can be
obtained. For instance, a distance function can be employed to represent some shapes for image segmentation \cite{kolmogorov04}
and surface segmentation \cite{boykov06a}.

\subsubsection{Max-Flow and Min-Cut}

A minimum cut is the cut that have the minimum cost called min-cut.
As an example, in foreground-background segmentation application, $V_s$ contains vertices
that corresponds to the foreground region in an image and $V_t$ contains vertices
that corresponds to the background region in an image.
We would like to to find a minimum cut containing two sets $V_s$ and $V_t$ such that
the foreground and the background regions can be identified.

We note that each edge can be interpreted as a pipe and its edge cost can be
considered as the capacity of this pipe. The max-flow problem is to find the largest
amount of flow allowed to pass from the source vertex to the sink vertex subject
to pipe capacity constraints and conservation of flows in the graph.
By the duality theorem \cite{ford62}, the max-flow problem is equivalent to the min-cut problem.
A globally optimum solution for min-cut can be found by using the max-flow algorithm (e.g., \cite{ford62,goldberg88}).
Other graph cut implementations include
push-relabel \cite{goldberg88} and pseudo-flows \cite{hochbaum98,juan06} techniques. They can be shown to be an iterative
algorithm by generating a sequence of cuts such that the sequence converges to a global optimum solution. These
iterative approaches can be interpreted as a splitting and merging method for finding an optimal graph partition.
Such efficient graph cut algorithms \cite{boykov01,boykov04}
are developed for image segmentation purpose.
Their numerical examples have shown that the performance of these algorithms is
significantly better than that of the standard max-flow technique. The main idea is to avoid
combinatorial computational and introduce an iterative approach for finding an optimal solution.
We will discuss this approach for solving Mumford-Shah segmentation model.

\subsubsection{Normalized Cuts}
In the literature, we know that a minimum cut may favour giving regions with a small number of vertices, see for instance
\cite{wu93,shi00}. To avoid such situation for partitioning out small regions, the use of the normalized cut is proposed by
Shi and Malik \cite{shi00}. The cost of a cut is defined as a fraction of the total edge connections to all the vertices
in the graph:
$$
{\rm cost}_n({\cal C}) = \frac{ {\rm cost}({\cal C}({\cal V}_s,{\cal V}_t)) }{
{\rm cost}({\cal C}({\cal V}_s,{\cal V})) } +
\frac{ {\rm cost}({\cal C}({\cal V}_s,{\cal V}_t)) }{
{\rm cost}({\cal C}({\cal V}_t,{\cal V})) },
$$
where ${\rm cost}({\cal C}({\cal V}_s,{\cal V}))$ is the sum of the cost of the edges between ${\cal V}_s$ and the whole set of
vertices and ${\rm cost}({\cal C}({\cal V}_s,{\cal V}))$ can be defined similarly.
It is shown in \cite{shi00} that the resulting optimization problem can be relaxed to solving an eigenvalue problem:
$$
( {\bf I} - {\bf D}^{-1/2} {\bf W} {\bf D}^{-1/2} ) {\bf y} = \lambda {\bf y}.
$$
The coefficient matrix is called a normalized Laplacian matrix. We note that spectral graph theory \cite{chung97} can be used
to study such normalized Laplacian matrix. The eigenvector corresponding to the second smallest eigenvalue of
normalized Laplacian matrix provides a normalized cut. The eigenvector corresponding to
the third smallest eigenvalue of normalized Laplacian matrix provides a partition of the first two regions identified by
the normalized cut. In practice, we can restart solving the partitioning problem on each subregion individually.

In the literature, other cuts are proposed and studied for image segmentation, for instance mean cut \cite{wang01},
ratio cut \cite{wang03} and ratio regions approach \cite{cox96}. In \cite{peng13}, some comparisons are presented
for different graph cut approaches. Recently,
an exact $l_1$ relaxation of the Cheeger ratio cut problem for multi-class transductive learning is studied in \cite{bresson14}.
In general, the problem of finding a cut (min-cut, normalized cut, ratio cut, mean cut and ratio region)
in an arbitrary graph is NP-hard. Definitely, efficient approximations to their solutions are required for image segmentation.

In some applications, a small number of pixels with known labels (foreground or background), the technique of random walks can be employed
to assign each pixel to the label for which the largest probability is calculated. The framework can be interpreted as discrete potential
theory an electrical circuits and the algorithm can be implemented on graphs that are constructed in Section II, see \cite{grady06,desquesnes13}.
A bilaterally constrained optimization model arising from the semi-supervised multiple-class image segmentation problem was developed 
in \cite{ng10,law12}. 

\subsection{The Mumford-Shah Model}

In \cite{boykov06}, Boykov et al. showed an interesting connection between
graph cuts and level sets \cite{sethian99}, and discussed how
combinatorial graph cuts algorithms can
be used for solving variational image segmentation problems
such as Mumford-Shah functionals \cite{mumford89}. In \cite{yuan10}, Yuan et al. further investigated novel max-flow and min-cut
models in the spatially continuous setting, and showed that the continuous max-flow models correspond to their respective
continuous min-cut models as primal and dual problems.

The Mumford-Shah model is an image segmentation model with a wide range of
applications in imaging sciences. Let $f$ be the target image. We would like to seek
a partition $\{ \Omega_i \}_{i=1}^{n}$ of the image domain $\Omega$,
and an approximation image $u$ which minimizes the functional
\begin{eqnarray} \label{mumford}
J( u, \{ \Gamma_i \}_{i=1}{n} )
& = &
\int_{\Omega} ( u - f )^2 dx + \beta \int_{\Omega \setminus \cup_i \Gamma_i} | \nabla u |^2 dx
\nonumber \\
& & + \nu \sum_{i=1}^{n} \int_{\Gamma_i} ds
\end{eqnarray}
where $\{ \Gamma_i \}_{i=1}^{n}$ denotes the interphases between the regions $\{ \Omega_i \}_{i=1}^n$.
It is interesting to note that when $u$ is assumed to be constant within each $\Omega_i$. The
second term in (\ref{mumford}) disappears and the resulting functional is given as follows:
\begin{equation} \label{mumford1}
J( u, \{ \Gamma_i \}_{i=1}^{n} )
= \int_{\Omega} ( u - f )^2 dx + \nu \sum_{i=1}^{n} \int_{\Gamma_i} ds,
\end{equation}
where
\begin{equation} \label{uc}
u = \sum_{i=1}^{n} c_i \xi_i
\end{equation}
and $\xi_i$ is the characteristic function of $\Omega_i$.

In \cite{chan00}, Chan and Vese proposed to use level set functions to represent the above functional and
solve the resulting optimization problem via the gradient descent method. Piecewise constant level set functions
are used in \cite{lie06}:
$$
\phi = i \quad {\rm in} \ \Omega_i, \ 1 \le i \le n.
$$
The relationship between the characteristic function and the level set function is given as follows:
$$
\xi_i = \frac{1}{\alpha_i} \prod_{j=1, j \ne i} ( \phi - j ) \quad {\rm with} \ \alpha_i = \prod_{k=1, k \ne i} (i-k).
$$
The length
term in (\ref{mumford1}) can be approximated by the total variation of the level set function itself.
The resulting Mumford-Shah functional becomes
\begin{equation} \label{mumford2}
J( u, \phi )
= \int_{\Omega} ( u - f )^2 dx + \nu \int_{\Omega} | \nabla \phi | dx.
\end{equation}
Many research works have been studied to minimize (\ref{mumford2}) by continuous optimization methods such
as the augmented Lagrangian method \cite{lie06,tai07} with the integer-valued constraint:
$\prod_{i=1}^{n} (\phi - i) = 0$.
Note that there are some variants of the total variation regularization term in the two-dimensional domain $(x_1,x_2)$ setting.
The isotropic form is given by $\int_{\Omega} \sqrt{ | \phi_{x_1}|^2 + | \phi_{x_2} |^2 } dx_1 dx_2$.
The anisotropic form is given by $\int_{\Omega} ( | \phi_{x_1}| + | \phi_{x_2} | ) dx_1 dx_2 $,
and its modified form is given by $\int_{\Omega} ( | \phi_{x_1}| + | \phi_{x_2} | + | R \phi_{x_1} | + | R \phi_{x_2} | ) dx_1 dx_2$,
where $R(\cdot)$ is the counterclockwise rotated gradient by $\pi/4$ radians used for creating more isotropic version.

\subsubsection{Discrete Models}
In \cite{bae09}, Bae et al. solved the minimization problem by graph cuts. They discretized the variational problem
(\ref{mumford2}) on a grid, and the discrete energy function can be written as follows:
\begin{equation} \label{mumford3}
J_d( u, \phi )
= \sum_{i} ( u_i - f_i )^2 + \nu \sum_{i} \sum_{j \in {\cal N}(i)} w_{i,j} | \phi_i - \phi_j |,
\end{equation}
where $i$ and $j$ refer to the grid points,
the weights $w_{i,j}$ are given by $w_{i,j} = \frac{1}{k \times distance(i,j)}$, $distance(i,j)$ is the distance between the two
grid points $i$ and $j$, and $k$ refers to the neighbourhood numbers in the discretization of different total variation forms.

\subsubsection{Graph Cuts Minimization}
For fixed values of $\{ c_i \}_{i=1}^{n}$, the minimizer of (\ref{mumford3}) can be solved by finding the minimum cut over a constructed
graph. It is not necessary to impose integer constraints in (\ref{mumford3}) to obtain integer-valued level set function $\phi_i$.
According to the optimization problem in (\ref{mumford3}), the set of vertices and the corresponding edges with their cost function
can be constructed suitably. The work on graph cuts for the two regions Mumford-Shah model can be found in \cite{darbon07,zehiry07}.

\begin{figure}
\centerline{\includegraphics[width =3.8in, height=2.2in]{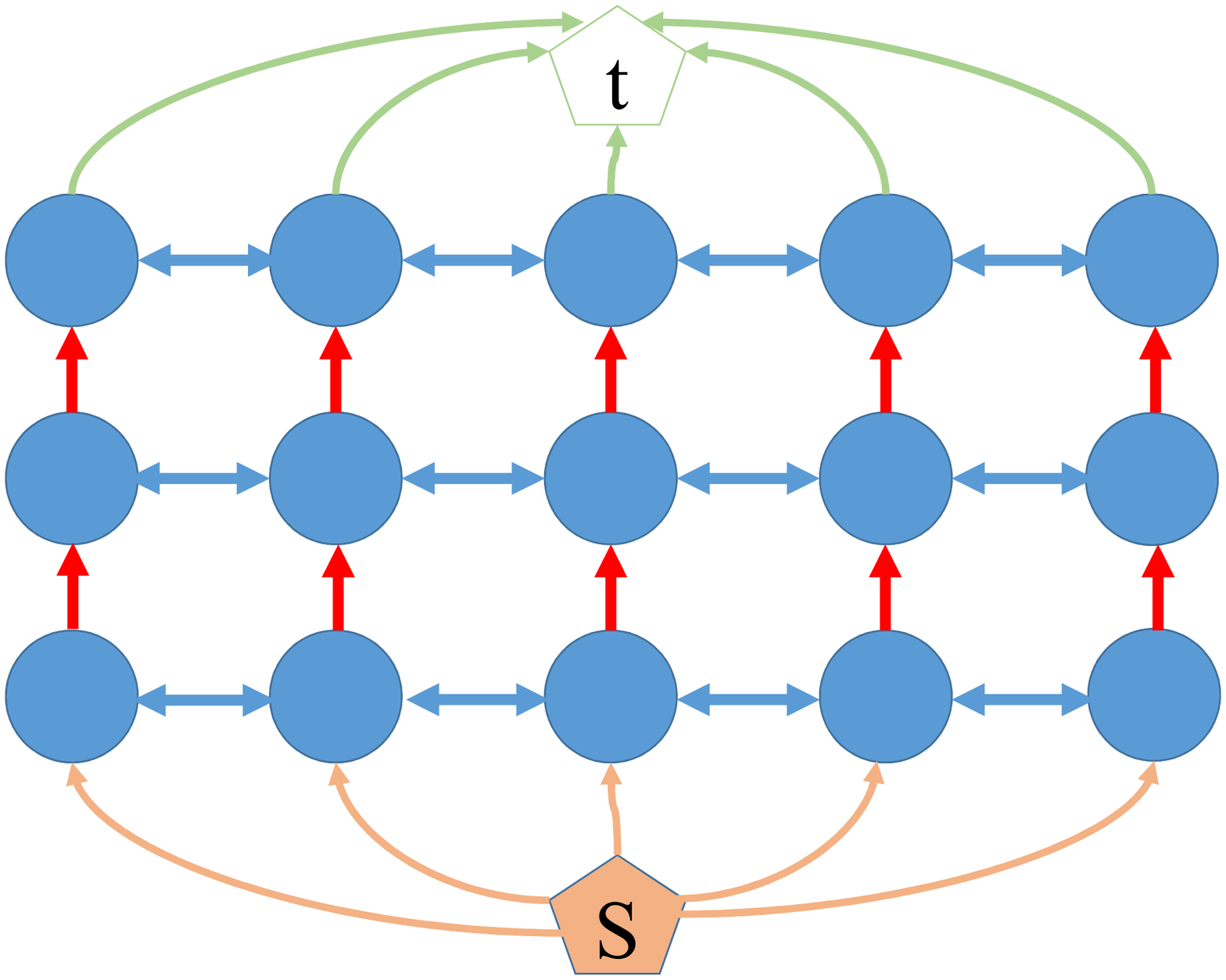}}
\vspace{-1cm}
\caption{A one-dimensional example of five grid points for three regions segmentation
(blue circle: image pixel; blue arrowed line: an edge between two grid point vertices; brown arrowed line: an
edge from the source vertex to a grid point vertex;
green arrowed line: an edge from a grid point vertex to the sink vertex; red arrowed line: an edge from one region to another region.}
\label{fig:gc1}
\end{figure}

For multiple regions, Bae et al. \cite{bae09} designed multiple layers to deal with multiple regions.
We refer to Figure\;\ref{fig:gc1} for one-dimensional example of five grid points and three regions segmentation for illustration.
The graph consists of three layers referring to three regions segmentation ($n=3$). Each layer contains five grid points as vertices (blue
circles).
The edges between grid points refer to their neighbourhoods (blue arrowed lines). The cost of these edges is related to the total variation regularization
term (or the boundary term for the discontinuity of the two neighbourhood grid points).
The source vertex and the sink vertex are also constructed in the graph. The cost of the
edges between the source vertex to the vertices
in the top layer, and between the sink vertex to the vertices in the bottom layer, refer to the region penalty term.
It was shown in \cite{bae09} that for any piecewise constant level set function $\phi$ taking values in $\{ 1,2,\cdots,n  \}$,
there exists a unique admissible cut on the constructed graph. The level set function $\phi$ corresponds to a minimum cut in
the constructed graph. After the level set function $\phi$ is determined, the values $\{ c_i \}_{i=1}^{n}$ can be minimized by
using the first term of (\ref{mumford3}), and they are given by
$$
c_i = \frac{ \sum_j f_j \xi_{i}(j) }{ \sum_j \xi_{i}(j) }, \quad i=1,2,\cdots,n.
$$
Numerical results in \cite{darbon07,zehiry07,bae09} have shown that this graph cut approach for solving the Mumford-Shah segmentation model
is superior in efficiency compared to the partial differential equation based approach.
Alternatively convex approaches to segmentation with active contours have also been considered, see for instance
\cite{lezoray12,drakopoulos12}.

This graph cut approach can be used to address a class of multi-labeling problems over a spatially continuous image domain, where
the data fitting term can be of any bounded function, see \cite{ishikawa09,yuan10,bae14}.
It can also be extended to the convex relaxation of Pott's model \cite{potts52} describing a partition of the continuous
domain into disjoint subdomains as the minimum of a weighted sum of data fitting and the length of the partition boundaries.
Recent research development along this direction include multi-class transductive learning
based on $L_1$ relaxations of Cheeger cut and
Mumford-Shah-Potts Model \cite{bresson14}, and image segmentation by using
the Ambrosio-Tortorelli functional and Discrete Calculus \cite{Foare16}.

\subsection{Graph BiLaplacian}

Graph Laplacian matrix plays a leading role in these graph-based optimization methods.
For example, Levin et al. \cite{levin04} proposed a semi-supervised image matting method with closed form solution.
Also Levin et al. \cite{levin08} proposed a spectral matting method based on the spectral analysis of
the matting Laplacian matrix derived in \cite{levin04}. Note that the matting Laplacian matrix can be viewed as a
generalization of the graph Laplacian.
Inspired by graph-based methods and their good performance, the graph Laplacian can be generalized to
its second-order graph Laplacian, namely graph biLaplacian \cite{fang2017}. In particular, in an image when a
vertex is only connected with its four neighbourhood vertices with equal edge weight, the graph biLaplacian
is a finite difference approximation to the biharmonic operator in a continuous setting.

The $i-$th component of graph Laplacian of $u \in \mathbb{R}^n$ is
$$
[\Delta_wu]_i = \sum_{j\in \mathcal{N}_i}w_{i,j}(u_i-u_j)
$$
where $\mathcal{N}_i$ denotes the neighbourhood of the $i$ vertex (all the vertices connected with the vertex $i$).
The $i$-th component of the graph biLaplacian
of $u$ can be considered as follows:
$$
(\Delta_w^2u)_i = \sum_{j\in \mathcal{N}_i}w_{i,j}([\Delta_w u]_i-[\Delta_w u]_j).
$$
The elements of the $i$-th row of the graph biLaplacian matrix $\Delta_w^2$ are:
\begin{eqnarray*}
(\Delta_w^2)_{i,i}&=& \sum_{j\in \mathcal{N}_i}w_{i,j}^2+\sum_{j\in \mathcal{N}_i}\sum_{l\in \mathcal{N}_i}w_{i,j}w_{i,l},\\
(\Delta_w^2)_{i,j}&=& -\sum_{k\in \mathcal{N}_j}w_{i,j}w_{j,k}-
\sum_{k\in \mathcal{N}_i}w_{i,j}w_{i,k}, \ j\in \mathcal{N}_i, \\
(\Delta_w^2)_{i,k}&=& \sum_{j\in \mathcal{N}_i}w_{i,j}w_{j,k}, \ k \in \mathcal{N}_j, \ j \in \mathcal{N}_i \ k \neq i.
\end{eqnarray*}
The normalized graph biLaplacian matrix can be defined similarly. The spectral properties of
graph biLaplacian and normalized graph biLaplacian can be found in \cite{fang2017}.

We remark that the above formulation of graph Laplacian and graph
biLaplacian is equivalent to the discretization of
the harmonic and biharmonic PDE equation with Neumann boundary condition respectively.
The harmonic equation is given by
\begin{equation*}
\begin{array}{l}\Delta u = 0, \mbox{in} \ \Omega, \quad \frac{\partial u}{\partial n}|_{\partial \Omega}=0.
\end{array}
\end{equation*}
The biharmonic equation is
\begin{equation*}
\begin{array}{l}\Delta^2 u = 0, \mbox{in} \ \Omega, \quad \frac{\partial u}{\partial n}|_{\partial \Omega}=0.
\end{array}
\end{equation*}
which comes from minimizing the following total squared curvature
\begin{equation*}
\min \int_\Omega |\Delta u|^2dx.
\end{equation*}
Harmonic and biharmonic equations and their numerical schemes
are widely studied and applied in data interpolation, computer vision
and image inpainting problems, see \cite{gaspar00,briggs74,bjostad83,hoffmann15,terzopoulos83,grady10}
and the references therein.

\section{Conclusion}
\label{sec:conclude}
Though graph signal processing (GSP) for large data networks has been studied intensively the last few years, applications of graph spectral techniques to image processing have received comparatively less attention. 
In this article, we overview recent developments of graph spectral algorithms for image compression, restoration, filtering and segmentation. 
Because a digital image lives naturally on a discrete 2D grid, one key challenge for graph-based image processing is the appropriate selection of the underlying graph that describes the image structure for the graph-based tools that operate on top. 
For compression, the description of the graph translates to side information coding overhead.
For restoration, filtering and segmentation, edge weights convey local signal similarity information, or a priori higher-level contextual information (e.g. saliency) that assist global processing operation.
For future work, one focus is to design application-specific graph structures that target specific tasks like image enhancement, while trading off performance with computation complexity.


\bibliographystyle{IEEEbib}
\bibliography{IEEEabrv,ref_gc,ref_em,ref_yt,ref_mn}

\end{document}